%% file: sample.tex
\documentclass[sigconf]{aamas} 

\usepackage{balance} 
\usepackage{amsmath}
\usepackage{multirow}
\usepackage{subcaption}
\usepackage[ruled,vlined,linesnumbered, resetcount]{algorithm2e}

\DeclareMathOperator*{\argmax}{arg\,max}

\usepackage{graphicx}

\setcopyright{ifaamas}
\acmConference[AAMAS '21]{Proc.\@ of the 20th International Conference on Autonomous Agents and Multiagent Systems (AAMAS 2021)}{May 3--7, 2021}{London, UK}{U.~Endriss, A.~Now\'{e}, F.~Dignum, A.~Lomuscio (eds.)}
\copyrightyear{2021}
\acmYear{2021}
\acmDOI{}
\acmPrice{}
\acmISBN{}

\acmSubmissionID{537}

\title{Reinforcement Learning for Unified Allocation and Patrolling in Signaling Games with Uncertainty }

\author{Aravind Venugopal$^{1}$ Elizabeth Bondi$^{2}$ Harshavardhan Kamarthi$^{3}$, Keval Dholakia$^{1}$,$^{3}$ Balaraman Ravindran$^{1}$ Milind Tambe$^{2}$}
\affiliation{$^1$Robert Bosch Centre for Data Science and AI, Indian Institute of Technology Madras\\
$^2$Center for Research on Computation and Society, Harvard University\\
$^3$College of Computing, Georgia Institute of Technology\\
}

\begin{abstract}
Green Security Games (GSGs) have been successfully used in the protection of valuable resources such as fisheries, forests and wildlife. While real-world deployment involves both resource allocation and subsequent coordinated patrolling with communication and real-time, uncertain information, previous game models do not fully address both of these stages simultaneously. Furthermore, adopting existing solution strategies is difficult since they do not scale well for larger, more complex variants of the game models.

We therefore first propose a novel GSG model that combines defender allocation, patrolling, real-time drone notification to human patrollers, and drones sending warning signals to attackers. The model further incorporates uncertainty for real-time decision-making within a team of drones and human patrollers. Second, we present CombSGPO, a novel and scalable algorithm based on reinforcement learning, to compute a defender strategy for this game model. CombSGPO performs policy search over a multi-dimensional, discrete action space to compute an allocation strategy that is best suited to a best-response patrolling strategy for the defender, learnt by training a multi-agent Deep Q-Network. We show via experiments that CombSGPO converges to better strategies and is more scalable than comparable approaches. Third, we provide a detailed analysis of the coordination and signaling behavior learnt by CombSGPO, showing group formation between defender resources and patrolling formations based on signaling and notifications between resources. \emph{Importantly, we find that strategic signaling emerges in the final learnt strategy}. Finally, we perform experiments to evaluate these strategies under different levels of uncertainty. 
\end{abstract}


\newcommand{\BibTeX}{\rm B\kern-.05em{\sc i\kern-.025em b}\kern-.08em\TeX}

\begin{document}


\pagestyle{fancy}
\fancyhead{}


\maketitle 


\input{intro}

\input{prelim}
\input{model}

\input{methodology}

\input{experiments}

\input{policies}

\input{discussions}






\bibliographystyle{ACM-Reference-Format} 
\bibliography{sample}


\end{document}


\maketitle

\section{Neural Network Architecture}

We first describe the neural network architecture for DDQNs used in the patrolling stage of the games. We then describe the neural networks that were used to represent allocation strategies for the defender and the attacker.

The DDQN had two convolution layers with the non-linear \textit{ReLU} activations between them. The first convolutional layer had 10 filters of size 3x3 and strides 1x1 while the second convolutional layer had 20 filters of size 3x3 and strides 1x1. The convolutional layers were followed by two fully-connected dense layers with 128 hidden units and 64 hidden units, respectively, with \textit{ReLU} activations in between. The last layer was a fully-connected dense layer with 5 units representing the 5 actions for the ranger DDQN and 15 units representing the actions for the drone DDQN.

Allocation strategies were represented by an actor network and a critic network. The actor network had a single fully-connected layer with a non-linear \textit{tanh} activation connected to two layers; the first being a fully-connected layer followed by a \textit{tanh} activation for predicting the mean of the action embedding distribution and the second, a fully-connected layer followed by a \textit{sigmoid} activation for predicting the variance of the action embedding distribution. The critic network had a fully-connected layer with \textit{tanh} activation followed by layer with a single unit for predicting the reward from choosing an allocation. Table 1 details the action embedding sizes used and Table 2 details the number of hidden units used in each layer for the networks described above.

\begin{table}[h]
\centering
\begin{tabular}{|l|l|l|l|}
\hline
Gridsize           & \#Attackers & Defender & Attacker           \\ \hline
\multirow{2}{*}{15x15}  & 1          & 50 & 2 \\ \cline{2-4} 
                         & 2          & 50 & 4 \\ \hline
\multirow{2}{*}{10x10} & 1          & 30 & 2 \\ \cline{2-4} 
                         & 2          & 30 & 4 \\ \hline
\end{tabular}
\caption{Action embedding sizes}
\label{tab:results1}
\end{table}

\begin{table}[h]
\centering
\begin{tabular}{|l|l|l|l|}
\hline
Gridsize           & \#Attackers & Defender(Actor and Critic) & Attacker(Actor and Critic)           \\ \hline
\multirow{2}{*}{15x15}  & 1          & 128 & 32 \\ \cline{2-4} 
                         & 2          & 128 & 32 \\ \hline
\multirow{2}{*}{10x10} & 1          & 64 & 32 \\ \cline{2-4} 
                         & 2          & 64 & 32 \\ \hline
\end{tabular}
\caption{Neural Network Architectures : Allocation Policies}
\label{tab:results1}
\end{table}

\section{Neural Network Training}

Tables 3 and 4 show the parameters used while training the ranger and drone DDQNs, respectively, for different grid sizes. The DDQNs follow an $\epsilon$ greedy policy and during training, $\epsilon$ = 1.0 initially and decays to 0.05 within 25000 steps. A discounting factor $\gamma$=0.99 is used for calculating the discounted rewards for DDQN training. For training, we use the Adam optimizer with $\beta1$ = 0.9 and $\beta2$ = 0.999.

While training the allocation policies, we use the Adam optimizer for updating the critic network, setting the critic learning rate at 1e-2 for both players. We use a batch size $n_{s}$=10 for all experiments. The actor networks are updated through competitive policy optimization. Table 5 shows the learning rates for competitive policy optimization, for experiments with different grid sizes and animal density distributions.

\begin{table}[h]
\centering
\begin{tabular}{|l|l|l|} \hline
 
 Hyperparameter & 15x15 & 10x10 \\  \hline

 Learning rate & 3e-4 & 3e-4\\ \hline
 Replay Buffer size & 1.92e5   & 1.92e4 \\ \hline
 Batch size & 32  & 32\\ \hline
 Target update step & 20 & 20\\\hline
 
 \hline
\end{tabular}
\caption{Parameters for Training : Drone DQN}
\label{table:2}
\end{table}

\begin{table}[h]
\centering
\begin{tabular}{|l|l|l|} \hline
 
 Hyperparameter & 15x15 & 10x10 \\  \hline

 Learning rate & 3e-4 & 3e-4\\ \hline
 Replay Buffer size & 0.64e5   & 0.64e4 \\ \hline
 Batch size & 32  & 32\\ \hline
 Target update step & 50 & 50\\\hline
 
 \hline
\end{tabular}
\caption{Parameters for Training : Ranger DQN}
\label{table:2}
\end{table}

\begin{table}[h]
\centering
\begin{tabular}{|l|l|l|}
\hline
Gridsize           & \#Attackers & Learning rate      \\ \hline
\multirow{2}{*}{15x15}  & 1           & 4e-5 \\ \cline{2-3} 
                         & 2          & 3e-5 \\ \hline
\multirow{2}{*}{10x10} & 1          & 3e-5 \\ \cline{2-3} 
                         & 2          & 3e-5 \\ \hline
\end{tabular}
\caption{Learning Rates : Allocation Actor Networks}
\label{tab:results1}
\end{table}

\newpage

\section{Animal Densities}

We perform experiments with two different kinds of animal densities : randomly distributed and spatially distributed. Here, we describe how we calculate spatially distributed animal densities. This distribution reflects the attractiveness of each cell in the park to a potential attacker. We assume that the distribution of targets that are of interest to attackers, is influenced by certain geographical and man-made features; namely : rivers, boundaries of the national park and roads (some of these features are used in (Gholami et al. 2018) to model attacker behavior). 

We consider boundaries at all edges of the grid and we also consider that a river and a road pass through the national park. We then rank each cell depending on it's distance from these features such that cells farther away have a higher rank, as shown in Fig. 4. We then take a weighted average of these ranks to get a measure of how attract a cell is to animals, with weights of 0.1, 0,1 and 0.8 given to the boundary, road and river ranks respectively and call it the animal rank. By further taking a weighted average of animal, river, road and boundary ranks with weights of 0.7, 0.05, 0.15 and 0.1, we arrive at the final animal density distribution.

\begin{figure}[h]
\centering
\includegraphics[width=.9\linewidth]{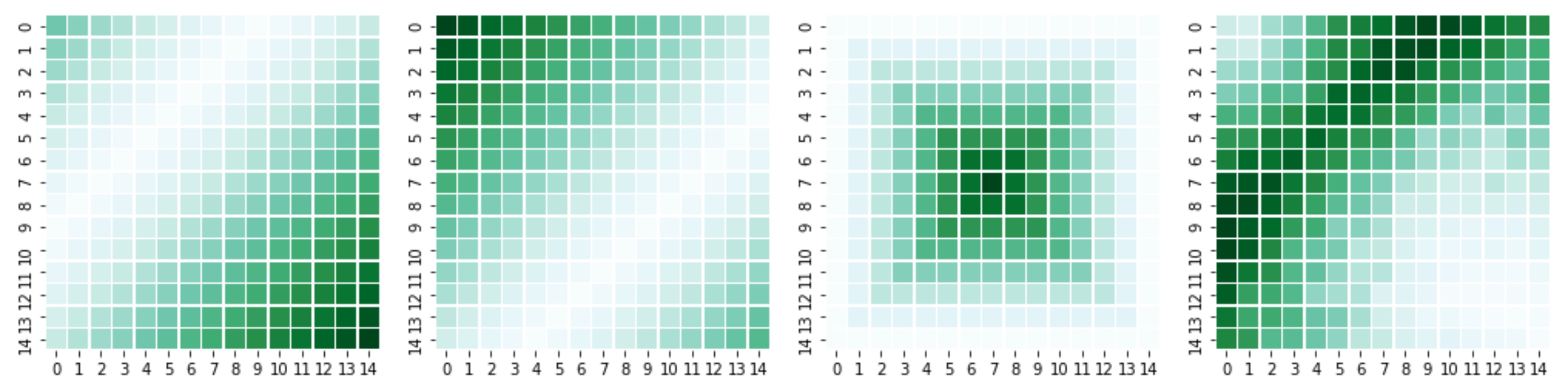}
\caption{Left to Right : Ranking of cells based on a. distance from river, b. distance from road, c. distance from boundaries and d. final animal density distribution for a 15x15 grid} 
\end{figure}

\clearpage
\section{Additional Experiments}
We show the learning curves for all baselines and CombSGPO under different uncertainty conditions in figures below, for gridsizes 15x15 and 10x10.

\begin{figure}[h]
    \centering
    \includegraphics[width=.7\linewidth]{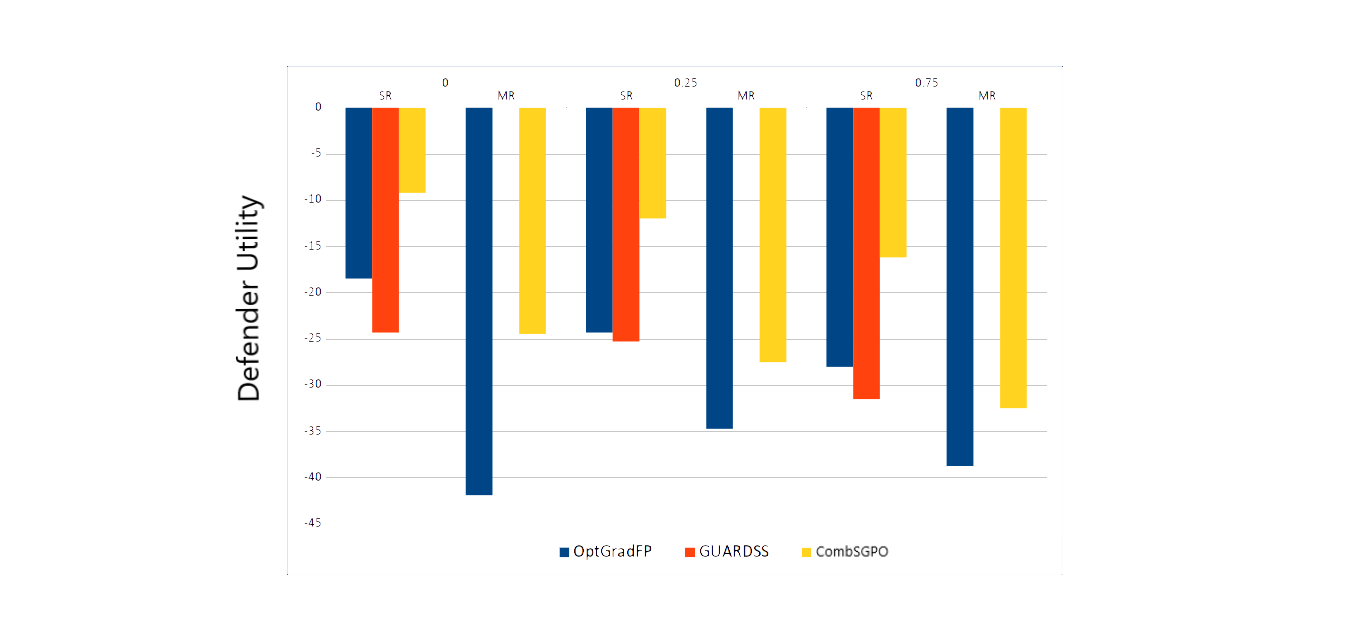}
    \caption{Average defender utilities for baselines and CombSGPO for $15\times 15$ grid with random animal density}
    \label{fig:rand}
\end{figure}

\begin{figure}[h]
    \centering
    \begin{subfigure}{85mm}
      \centering
      \includegraphics[width=95mm]{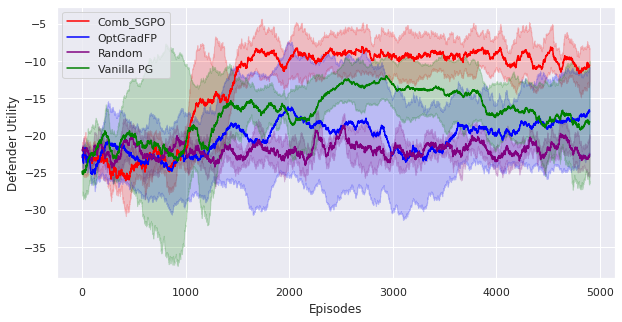}
      \caption{SR}
    \end{subfigure}
    \begin{subfigure}{85mm}
      \centering
      \includegraphics[width=95mm]{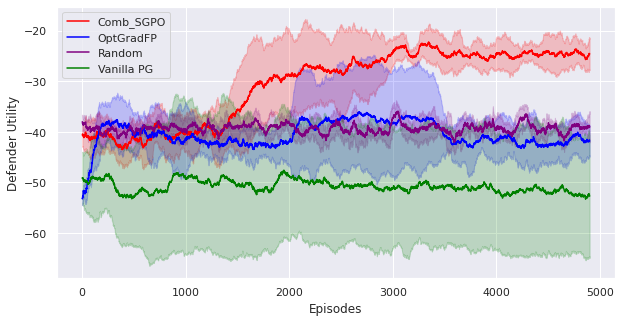}
      \caption{MR}
    \end{subfigure}
   \caption{Training curves for experiments on a 15x15 grid ($\beta$=0,$\kappa$=0)}
   \label{fig:results1}
\end{figure}

\begin{figure}[h]
    \centering
    \begin{subfigure}{.5\linewidth}
      \centering
      \includegraphics[width=\linewidth]{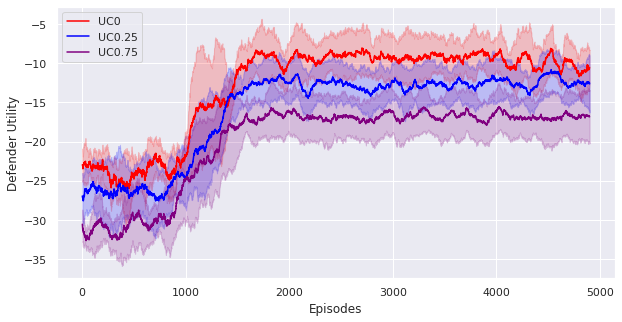}
      \caption{SR}
    \end{subfigure}%
    \begin{subfigure}{.5\linewidth}
      \centering
      \includegraphics[width=\linewidth]{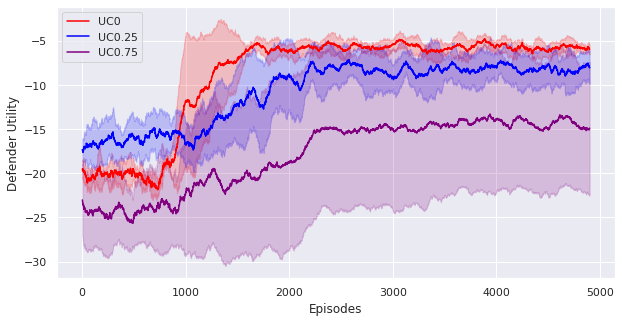}
      \caption{SS}
    \end{subfigure}
    \begin{subfigure}{.5\linewidth}
      \centering
      \includegraphics[width=\linewidth]{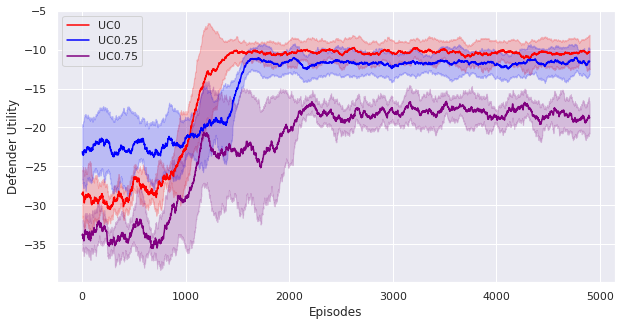}
      \caption{MS}
    \end{subfigure}%
    \begin{subfigure}{.5\linewidth}
      \centering
      \includegraphics[width=\linewidth]{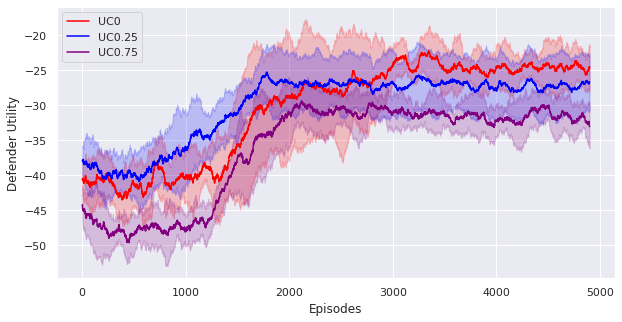}
      \caption{MR}
    \end{subfigure}
   \caption{Training reward curves for CombSGPO with different uncertainties for $15\times 15$ grid}
   \label{fig:coPGall}
\end{figure}

\begin{figure}[h]
    \centering
    \begin{subfigure}{.5\linewidth}
      \centering
      \includegraphics[width=\linewidth]{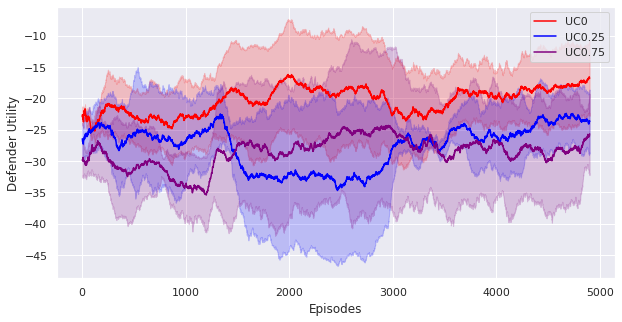}
      \caption{SR}
    \end{subfigure}%
    \begin{subfigure}{.5\linewidth}
      \centering
      \includegraphics[width=\linewidth]{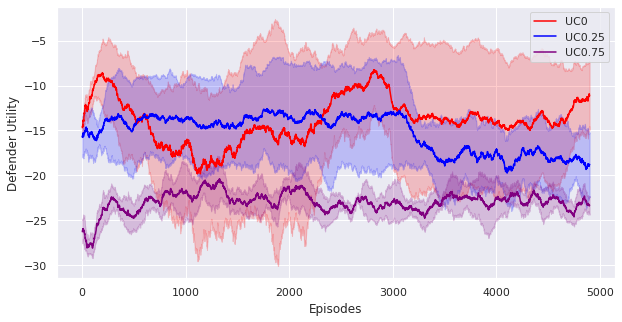}
      \caption{SS}
    \end{subfigure}
    \begin{subfigure}{.5\linewidth}
      \centering
      \includegraphics[width=\linewidth]{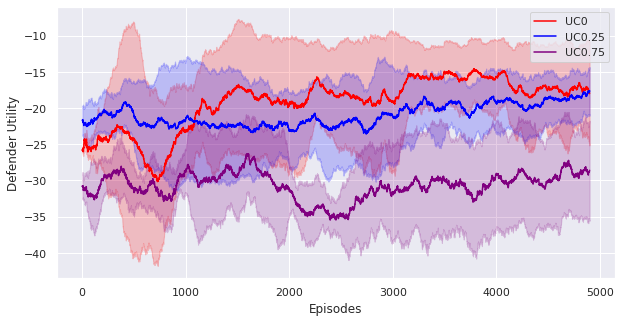}
      \caption{MS}
    \end{subfigure}%
    \begin{subfigure}{.5\linewidth}
      \centering
      \includegraphics[width=\linewidth]{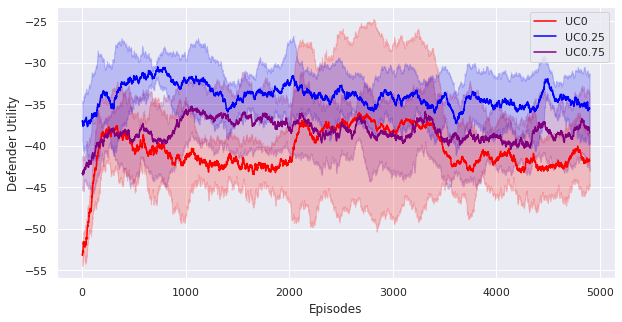}
      \caption{MR}
    \end{subfigure}
   \caption{Training reward curves for OPTGradFP with different uncertainties for $15\times 15$ grid}
   \label{fig:optgradall}
\end{figure}


\begin{figure}[h]
    \centering
    \includegraphics[width=\linewidth]{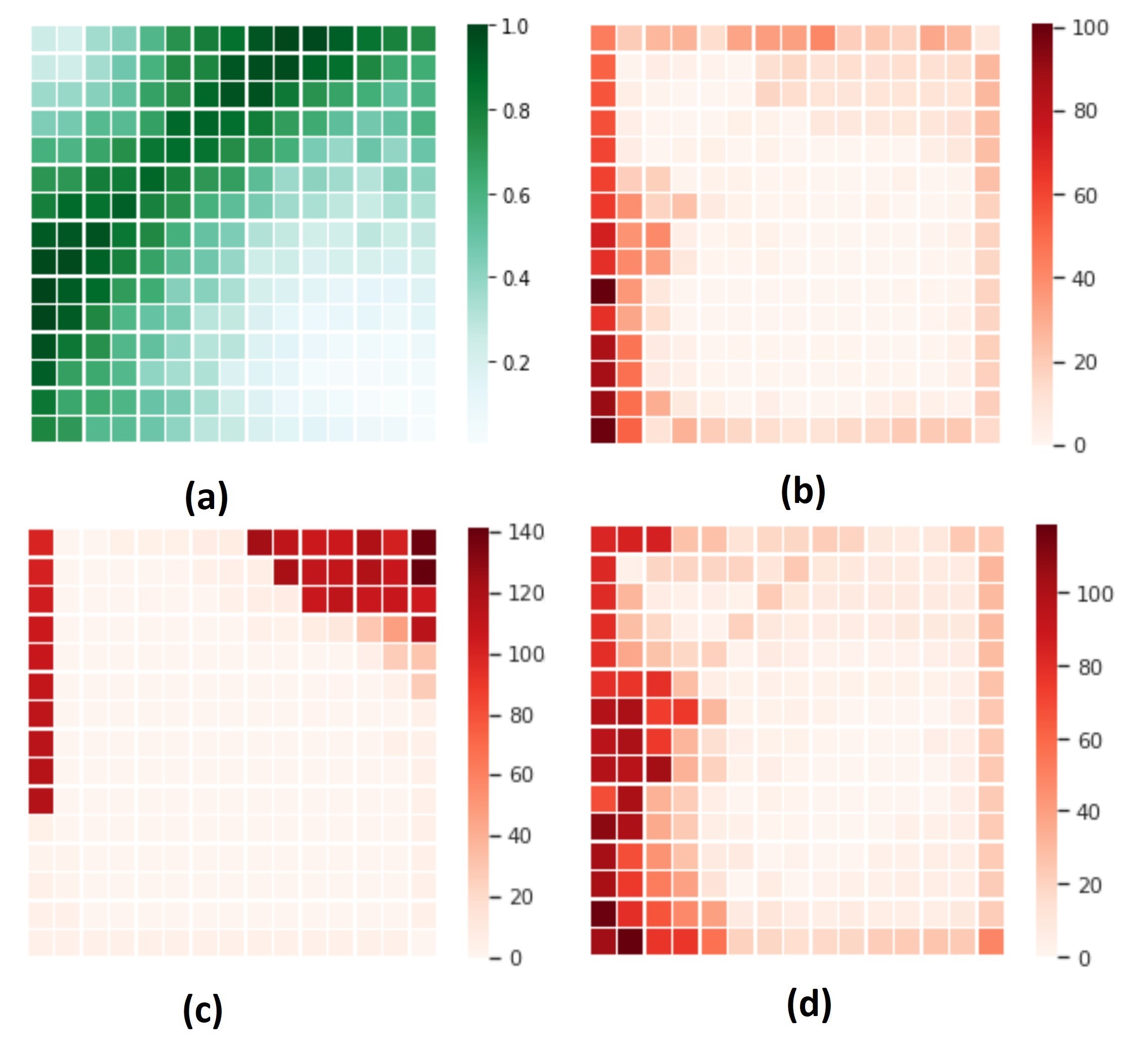}
    \caption{Visualization of sampled attacks (heatmaps of 100 sampled episodes) for MS setting for 15x15 gridsize. (a) shows the animal density distribution while the other figures show attacker heatmaps for: (b) CombSGPO, (c) OptGradFP, (d) PG}
    \label{fig:uc}
\end{figure}

\clearpage

\begin{figure}[h]
    \centering
    \includegraphics[width=\linewidth]{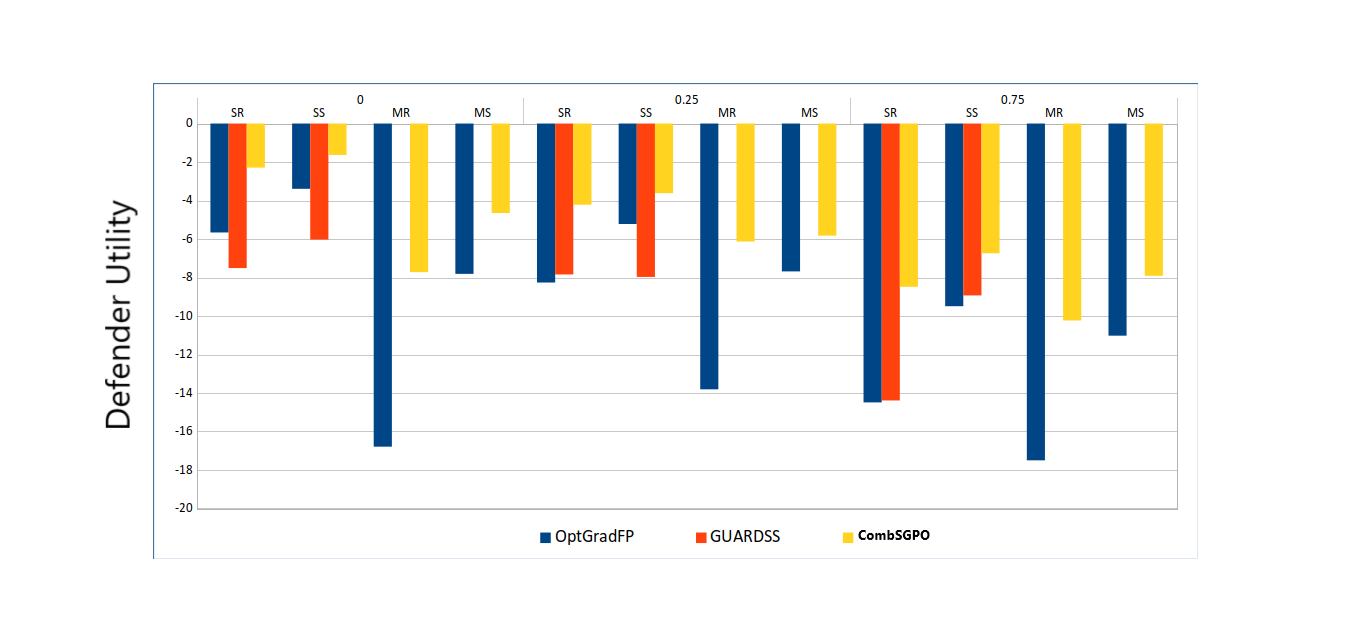}
    \caption{Average utilities for baselines and CombSGPO for $10\times 10$ grid}
    \label{fig:10x10}
\end{figure}

We also report the results for $10\times 10$ grid size which was used in GUARDSS paper. We observe from figure \ref{fig:10x10} that our model clearly outperforms OptGradFP and GUARDSS in all cases.

\begin{figure}[h]
    \centering
    \begin{subfigure}{.5\linewidth}
      \centering
      \includegraphics[width=\linewidth]{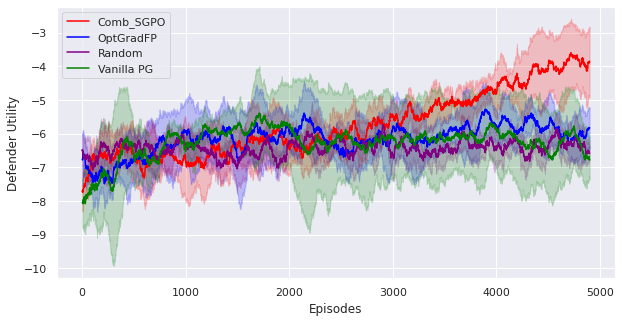}
      \caption{SR}
    \end{subfigure}%
    \begin{subfigure}{.5\linewidth}
      \centering
      \includegraphics[width=\linewidth]{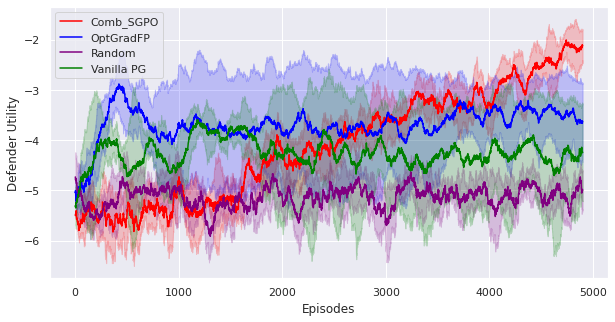}
      \caption{SS}
    \end{subfigure}
    \begin{subfigure}{.5\linewidth}
      \centering
      \includegraphics[width=\linewidth]{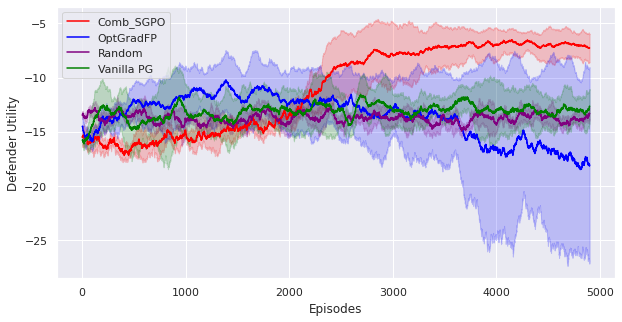}
      \caption{MR}
    \end{subfigure}%
    \begin{subfigure}{.5\linewidth}
      \centering
      \includegraphics[width=\linewidth]{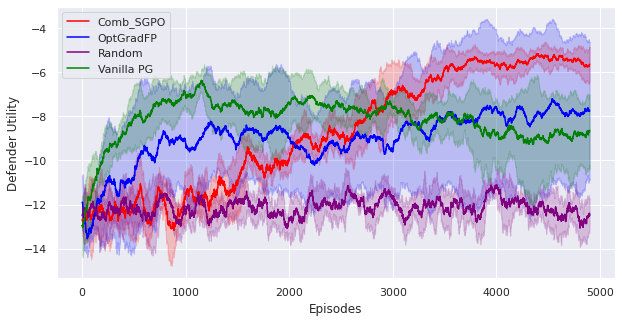}
      \caption{MS}
    \end{subfigure}
   \caption{Training reward curves (5000 episodes) comparing CombSGPO against  baselines ($\beta$=0,$\kappa$=0) for $10\times 10$ grid}
   \label{fig:coPGall10}
\end{figure}

\begin{figure}[h]
    \centering
    \begin{subfigure}{.5\linewidth}
      \centering
      \includegraphics[width=\linewidth]{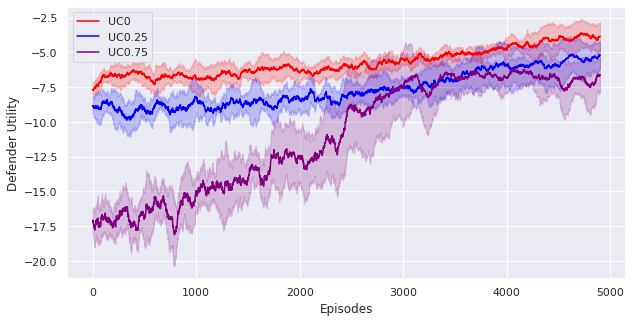}
      \caption{SR}
    \end{subfigure}%
    \begin{subfigure}{.5\linewidth}
      \centering
      \includegraphics[width=\linewidth]{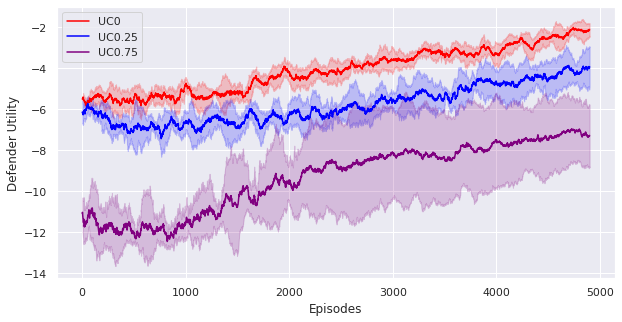}
      \caption{SS}
    \end{subfigure}
    \begin{subfigure}{.5\linewidth}
      \centering
      \includegraphics[width=\linewidth]{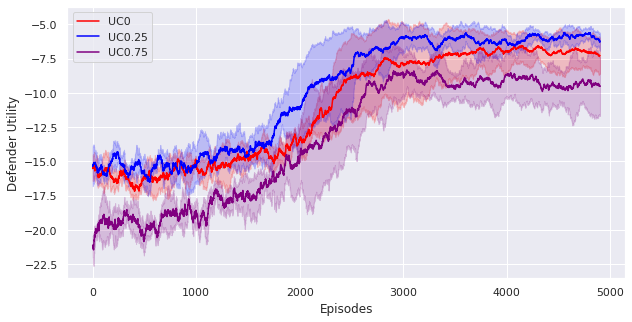}
      \caption{MR}
    \end{subfigure}%
    \begin{subfigure}{.5\linewidth}
      \centering
      \includegraphics[width=\linewidth]{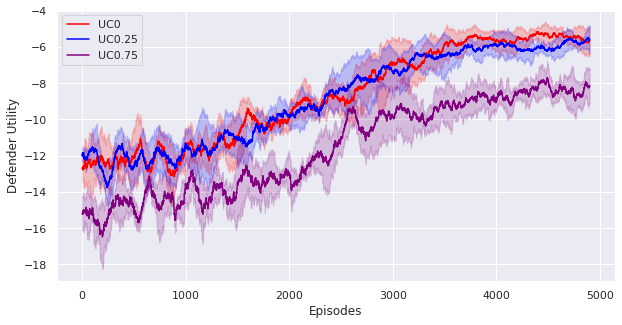}
      \caption{MS}
    \end{subfigure}
   \caption{Training reward curves (5000 episodes) for CombSGPO with different uncertainties for $10\times 10$ grid}
   \label{fig:coPGall10}
\end{figure}

\begin{figure}[h]
    \centering
    \begin{subfigure}{.5\linewidth}
      \centering
      \includegraphics[width=\linewidth]{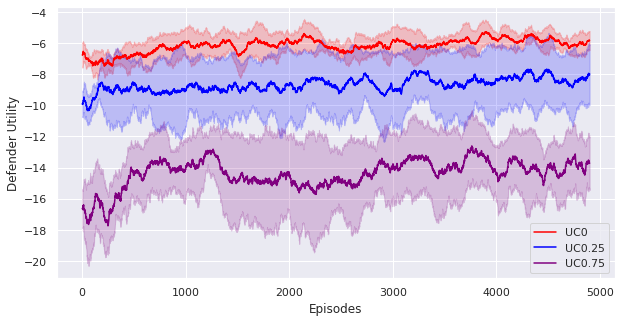}
      \caption{SR}
    \end{subfigure}%
    \begin{subfigure}{.5\linewidth}
      \centering
      \includegraphics[width=\linewidth]{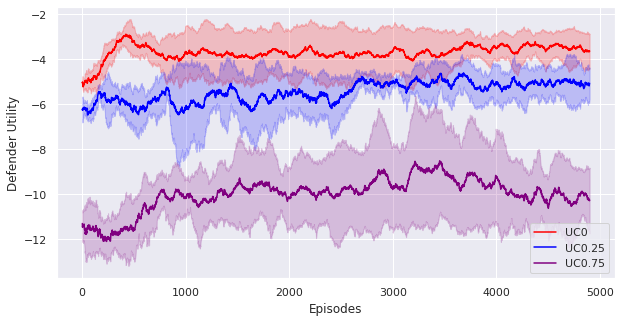}
      \caption{SS}
    \end{subfigure}
    \begin{subfigure}{.5\linewidth}
      \centering
      \includegraphics[width=\linewidth]{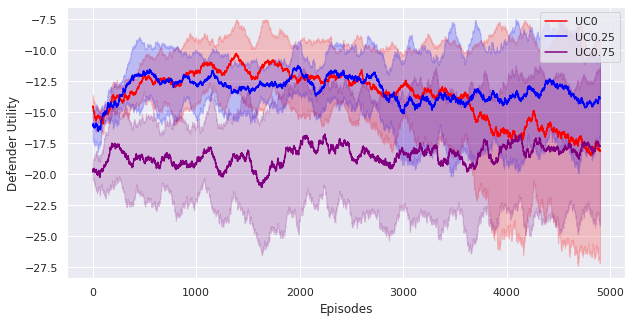}
      \caption{MR}
    \end{subfigure}%
    \begin{subfigure}{.5\linewidth}
      \centering
      \includegraphics[width=\linewidth]{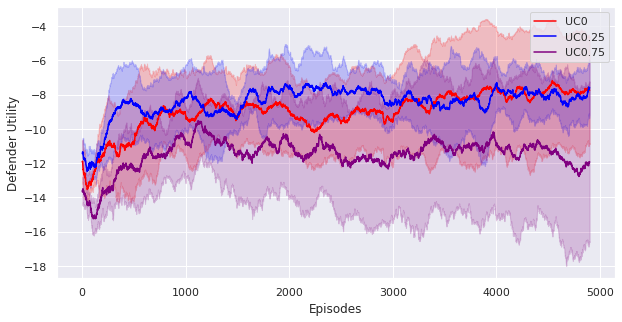}
      \caption{MS}
    \end{subfigure}
   \caption{Training reward curves(5000 episodes) for OPTGradFP with different uncertainties for $10\times 10$ grid}
   \label{fig:optgradall10}
\end{figure}


%% file: intro.tex
\section{Introduction}

Conservation of natural areas, forests, fisheries, and wildlife is increasingly important in the face of climate change and biodiversity loss. In order to model the protection of these resources, researchers have turned to game theory. In particular, Stackelberg Security Games have found extensive use in the development and deployment of complex systems for resource allocation and patrolling in domains such as airport security, transportation and protection of critical infrastructure\cite{krausarmor,tsai2009iris,yin2012trusts,pita2008deployed}. Green Security Games (GSGs), a sub-class of SSGs, have been used to model repeated interactions between defenders and attackers in green domains such as forests, fisheries and wildlife with the aim to assist real-world security agencies against illegal logging, illegal fishing and wildlife poaching. The defenders protect a finite set of targets with limited resources while attackers plan attacks with knowledge about the defender's strategy.  Traditional approaches to compute optimal defender strategies for resource allocation or patrolling rely on linear programming (LP) and mixed integer linear programming (MILP) \cite{fang2015security,fang2016deploying,xu2017optimal}. While a number of such approaches have been developed for GSGs with real-time information \cite{bondi2020signal,xu2018strategic,basilico2015security}, these techniques do not scale well with large complex GSGs such as GSG-I \cite{wang2019deep}.

Recently, there has been increasing interest in using Deep Reinforcement Learning algorithms for computing optimal defender strategies for GSGs \cite{wang2019deep, kamra2018policy, kamra2019deepfp}. \cite{kamra2018policy} and \cite{kamra2019deepfp} focus only on resource allocation and not on computing patrolling strategies. The focus of \cite{wang2019deep} is only on computing patrolling strategies in a game model that involves a single defender and a single attacker. Neither of these game models consider uncertainty in real-time observation information of the agents or on learning signaling strategies for communication between multiple defender agents. Although \cite{bondi2020signal} also incorporates allocation and signaling recommendations for a team of defender resources that comprises of sensors and human patrollers with uncertain, real-time information, it does not scale effectively to larger parks, multiple attackers, patrols, etc. However, real-world scenarios often involve allocation as well as patrolling with defender resources in large areas in the presence of uncertainty in scenarios that benefit from communication and coordination between resources during patrols. 

We therefore introduce a unified approach for computing strategies for resource allocation and subsequent patrolling in GSGs while also taking into account uncertainty in real-time information, to develop strategic defender allocation and patrolling strategies with inter-resource communication and signaling. We list our contributions as follows:

\begin{enumerate}
\item We propose a game model with two stages that captures real-time information through signaling and communication within a team of players; the first stage for handling resource allocation and the second stage for patrolling once the allocation decision has been made. This will be an imperfect-information Extensive Form Game (EFG) capable of representing real-world scenarios with multiple types of defender agents and multiple attackers.

\item We present CombSGPO (\textbf{C}ombined \textbf{S}ecurity \textbf{G}ame \textbf{P}olicy \textbf{O}ptimization), a novel solution strategy combining resource allocation with patrolling in extensive form GSGs by integrating recent advances in competitive policy optimization \cite{prajapat2020competitive} and multi-agent reinforcement learning.
We also provide extensive experimental results showing the effectiveness of our solution strategy in different types of environments and in the presence of different levels of uncertainty. We show that our approach ensures very fast convergence to better strategies when compared with OptGradFP \cite{kamra2018policy}, a fictitious play based approach and with GUARDSS \cite{bondi2020signal}, a MILP-based approach for computing optimal defender strategies.

\item Finally, we analyze the  signaling behaviors learnt by CombSGPO, illustrating group formation and communication among defender resources, including signaling and notifications between resources for apprehending attackers. We find that strategic signaling in fact emerges in the final learnt strategy, including a behavior in which drones start signaling or notifying more if they do not see attackers for some time. 
\end{enumerate}


%% file: prelim.tex
\section{Preliminaries}

\subsection{Green Security Games}

GSGs \cite{basak2016combining, fang2015security} belong to the larger class of SSGs \cite{tambe2011security, korzhyk2011stackelberg}, which feature interactions between two players, a leader and a follower. The two players need not necessarily be individuals. They could also represent multiple agents executing a joint strategy towards a common utility-maximizing goal. A pure strategy or policy is a deterministic mapping from a player's observation space to its strategy space. Throughout this paper, we use the words strategy and policy interchangeably. Each player can either play a pure strategy or a mixed strategy which is a probability distribution over pure strategies. The leader first commits to a strategy and the follower then optimizes its reward, given the leader's strategy. 

GSGs are essentially SSGs applied to green security domains and feature repeated interactions between a defender, which is the leader, and an attacker, which is the follower. With limited resources, the defender has to protect a set of targets under repeated attack from the attackers. 




\subsection{Competitive Policy Optimization}
In a zero-sum game with two players, the policy gradient algorithm \cite{sutton2000policy} derives policy updates for each agent by maximizing (or minimizing) the linear approximation of the game objective. This does not take the interaction between players into account and therefore, updates the policy of each agent assuming that the other agent is stationary. This generally leads to poor results and non-convergence to Nash Equilibrium (NE) in two-player, zero-sum games.

In contrast, Competitive Policy Optimization (coPO) \cite{prajapat2020competitive} derives policy updates for each agent by computing the NE of the bilinear approximation of the game objective as follows:
\begin{equation}
\theta^{1} \leftarrow \theta^{1} +  \argmax_{\Delta\theta^{1}:\Delta\theta^{1}+\theta^{1}}{\Delta\theta^{1}}^{T} D_{\theta^{1}}\eta  +  {\Delta\theta^{1}}^{T} D_{\theta^{1}\theta^{2}}\eta\Delta\theta^{2} - \frac{1}{2\alpha}\parallel\Delta\theta^{1}\parallel^{2}
\end{equation}
\begin{equation}
\theta^{2} \leftarrow \theta^{2} + \argmax_{\Delta\theta^{2}:\Delta\theta^{2}+\theta^{2}}{\Delta\theta^{2}}^{T} D_{\theta^{2}}\eta + {\Delta\theta^{2}}^{T} D_{\theta^{2}\theta^{1}}\eta\Delta\theta^{1} - \frac{1}{2\alpha}\parallel\Delta\theta^{2}\parallel^{2}
\end{equation}

\noindent where $\eta(\theta^{1},\theta^{2})$ represents the game objective  and $\theta^{1}$ and $\theta^{2}$ parameterize the policies of player 1 and player 2, respectively. $\alpha$ represents the stepsize. As a result of these updates, each agent updates its policy considering what its opponent's move will be at the current timestep and in the future. In a two-player, zero-sum game, this greatly reduces non-stationarity in the environment and leads to stable convergence to NE.



%% file: model.tex
\section{Game Model}

\begin{figure}[bp]
\centering
\includegraphics[width=80mm]{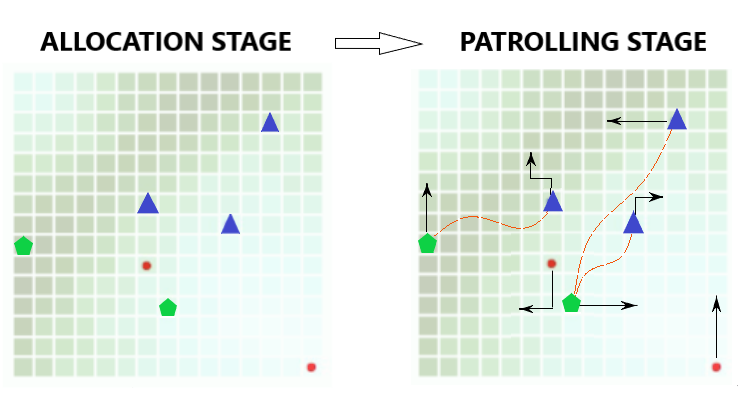}
\caption{Game model showing the allocation and patrolling stages. Rangers are represented by green pentagons, drones by blue triangles and attackers by red dots. Dotted orange lines show communication between drones and rangers. The animal density distribution is shown in the background.}
\end{figure}

We represent our environment by a gridworld, which represents a protected area/park to be conserved, e.g., protected from illegal poaching. Each cell in the gridworld represents a region in this park with a certain natural resource, e.g., an animal density. 
We will use animal density throughout the remainder of the paper, but this could also apply to tree density in the case of logging, for example. Details about animal density calculation are provided in Supplementary Material.

The defender has $n$ resources of which $n_{r}$ are park rangers (human patrollers) and $n_{d}$ are surveillance drones, while there are $n_{a}$ attackers. 
We collectively refer to the drones, rangers and attackers as agents.

The game is played out in two stages (illustrated in Fig. 1):
\begin{enumerate}
    \item \textbf{Allocation Stage} : The defender allocates $n_d$ drones and $n_r$ rangers to different cells in the environment, which represent the initial locations from which they should start patrolling. In response, the attacker allocates $n_a$ attackers to their initial target cells from which to start attacking.
    \item \textbf{Patrolling Stage} : The defender agents (drones and rangers) execute a coordinated patrolling strategy to protect potential attacker targets while the attacker agents move from one target to the next, causing damage to the park. The patrolling stage ends and the game is terminated when the attacker flees the park, is captured, or when the time exceeds $T$ steps, whichever comes first.
\end{enumerate}

We now describe the details of the patrolling stage. Each drone is equipped with an object detector \cite{bondi2018spot} that detects an attacker if it is in the current cell of the drone. Drones cannot apprehend an attacker but can either notify a park ranger of the detection and current position, or send a warning signal to an attacker (e.g., turning on lights) to deter an attack and lead him to flee. An attacker does not do damage to the cells he visits while fleeing, and has fled successfully once he reaches a cell on any edge of the park. While an attacker can observe a drone signal, he cannot observe notifications sent by drones to rangers. Rangers can apprehend an attacker if they are in the same cell, including while the attacker flees. 

We incorporate uncertainty in real-time information during the patrolling stage. We focus on two types of uncertainties: detection and observation. Detection uncertainty is uncertainty in whether an attacker has been detected by a drone in the same cell, e.g., due to occlusion by trees or due to misdetection of human-like objects as attackers. This makes it difficult for the rangers to make decisions based on signals received from drones. As in \cite{bondi2020signal}, we consider false negative detections specifically, which we denote by $\beta$. 
Observation uncertainty is on the side of the attacker, where a drone signals to the attacker but the attacker might not see it, e.g., lights and signals may be blocked by trees. As a result, an attacker may not behave as expected and might not be deterred even by signals from drones. We denote this uncertainty, which is due to false negative observations by the attacker, by $\kappa$.

Finally, during patrolling, each agent can only observe the current cell that it is in and thus, has only local observations. We also assume that each agent has memory to keep track of its observations from the start of a game. The game progresses in discrete timesteps. An agent can move only to any one of its neighbouring cells at each timestep. Action spaces for the agents are as follows:
\begin{itemize}
    \item Drone ($A_{d}$) : \textit{[ up, down, left, right, stay ] x [ Signal Attacker, Notify Ranger, NoOp ]}
    \item Ranger ($A_{r}$) : \textit{[ up, down, left, right, stay ]}
    \item Attacker ($A_{a}$): \textit{[ up, down, left, right, stay ]}
\end{itemize}


Now, we describe the rewards. At each timestep in the patrolling stage,  defender agents collectively receive a positive reward $r^{+}$ if an attacker is caught, and a zero reward if the attacker flees. If the attacker is not caught or deterred in the current timestep, the defender agents receive a negative reward $r^{-}$ for each target attacked by each attacker, which is proportional to the animal density of the attacked target. In addition to this, a drone agent can receive $r^{c}$ for notifying or signaling when an attacker is detected by it and $\bar{r^{c}}$ for notifying or signaling without detecting an attacker. This ensures efficient communication by penalizing false and redundant notifications or signaling. The sum of rewards of all defender agents at any timestep $t$ is given by $r^{def}_{t}$.

In the allocation stage, the payoff to the defender from choosing an allocation is $R^{d}$, given by $\sum_{t=0}^{T}$ $r^{def}_{t}$, which is the cumulative reward from playing out the patrolling stage from that allocation. When the attacker chooses an allocation, it receives a payoff of $R^{a}$, where $R^{a}$ = -$R^{d}$.

%% file: methodology.tex
\section{Methodology} 


We introduce our algorithm, CombSGPO, designed for computing the optimal defender strategy for the two-stage game model introduced in the previous section. CombSGPO first computes a defender patrolling strategy for the patrolling stage of our game and then chooses defender allocations that are best suited for this patrolling strategy. In this section, we first describe the behavioral model we use for our attacker. We then describe how we compute defender strategies for the patrolling and allocation stages and introduce CombSGPO, which combines the patrolling and allocation strategy computations. Throughout this section, 'allocation' refers to the cells to which defender or attacker agents are deployed during the allocation stage.

We denote defender and attacker allocation strategies by $\pi^{d}$ and $\pi^{a}$ respectively. Defender and attacker strategies for the patrolling stage are denoted by $\phi^{d}$ and $\phi^{a}$ respectively.


\begin{figure*}
  \includegraphics[width=140mm]{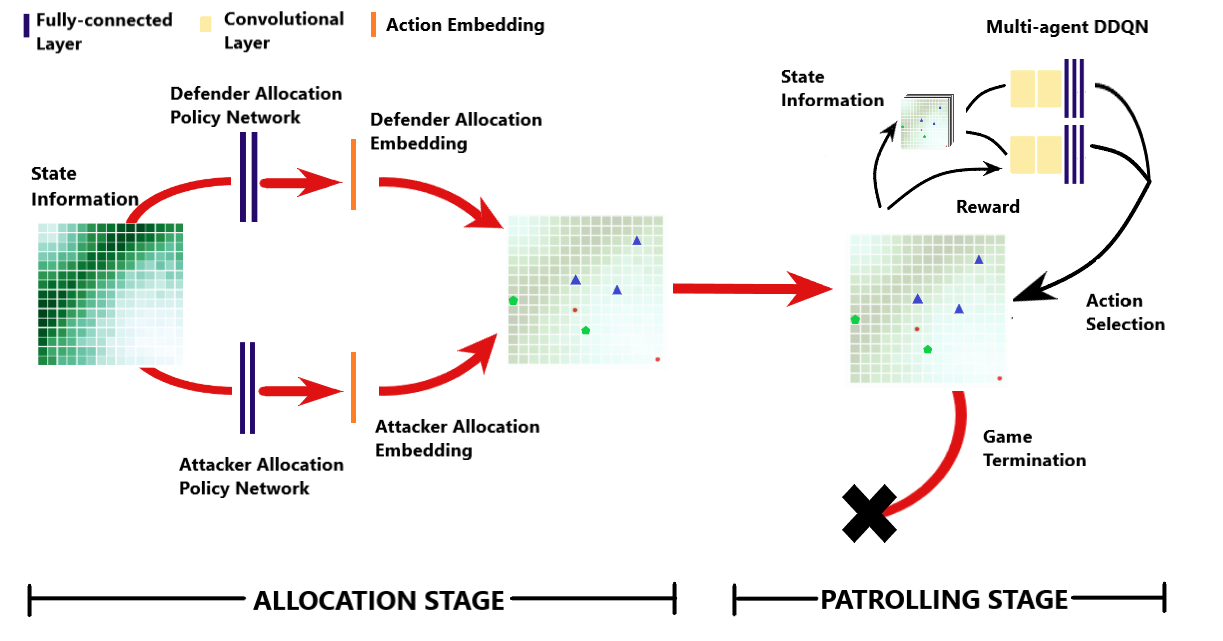}
  \caption{CombSGPO algorithm. (The state information in the allocation stage represents the animal density distribution. Green dots: Rangers, Blue dots: Drones, Red dots: Attackers}
\end{figure*}

\subsection{Attacker Behavioral Model} 
The attacker allocation strategy, $\pi^{a}$, is parameterized by the weights of a neural network which are updated to maximize the best response against the defender's allocation strategy. The attacker strategy for the patrolling stage, $\phi^{a}$, is a heuristic strategy which is based on knowledge of the animal density of each cell in the park as well as the allocations of the defender agents.

During the patrolling stage, the attacker's choice of target cell depends on the cell's animal density as well as its distance from defender agent allocations. This is done to reflect that attackers will try to attack regions with higher numbers of animals while trying to be as distant from the defender agents as possible. Attackers rank each cell based on its minimum distance from the  allocation of any defender agent, such that cells that are farther away from defender agents are given higher ranks. They then assign a score to each cell, equal to the average of the animal density of that cell and the normalized distance rank given to that cell. Over subsequent episodes, the score for a cell is updated as follows:
\begin{equation}
s_{av} = 0.5ad + 0.5dr 
\end{equation}
\begin{equation}
score_{ep+1} = score_{ep} + 0.1(s_{av} - score_{ep})
\end{equation}
where $ad$, $dr$ and $s_{av}$ are the animal density, normalized distance rank and average score of each cell. At each timestep in an episode, the neighbouring cell with the highest score is chosen by an attacker as the next target to attack.

\subsection{Computing Patrolling Strategy}

We formulate the patrolling stage as a Partially Observable Markov Decision Process (POMDP) consisting of $S$; 
the set of states, $A\rightarrow A_{r} \times A_{d}$ ; the set of all possible actions, $\mathcal{O} \rightarrow \mathcal{O}_{r} \times \mathcal{O}_{d}$ ; 
the set of all observations where $\mathcal{O}_{r} \rightarrow \mathcal{O}^{1}_{r}$ x ...x $\mathcal{O}^{n_r}_{r}$ is the set of observations of all rangers and $\mathcal{O}_{d}\rightarrow \mathcal{O}^{1}_{d}$ x ...x $\mathcal{O}^{n_d}_{d}$ is the set of observations of all drones, $\mathcal{T}$ $\rightarrow$ $S$ x $A$ ; the state transition function, $R$ : $S$ × $A$ → $\mathcal{R}$ ; the reward function and discount factor $\gamma \in$  [0, 1].

To compute coordinated patrolling strategies with communication for a team of $n_{r}$ rangers and $n_{d}$ drones, we use a multi-agent Double Deep Q Network (DDQN) \cite{van2015deep} .

\begin{figure}[bp]
    \centering
    \includegraphics[width=\linewidth]{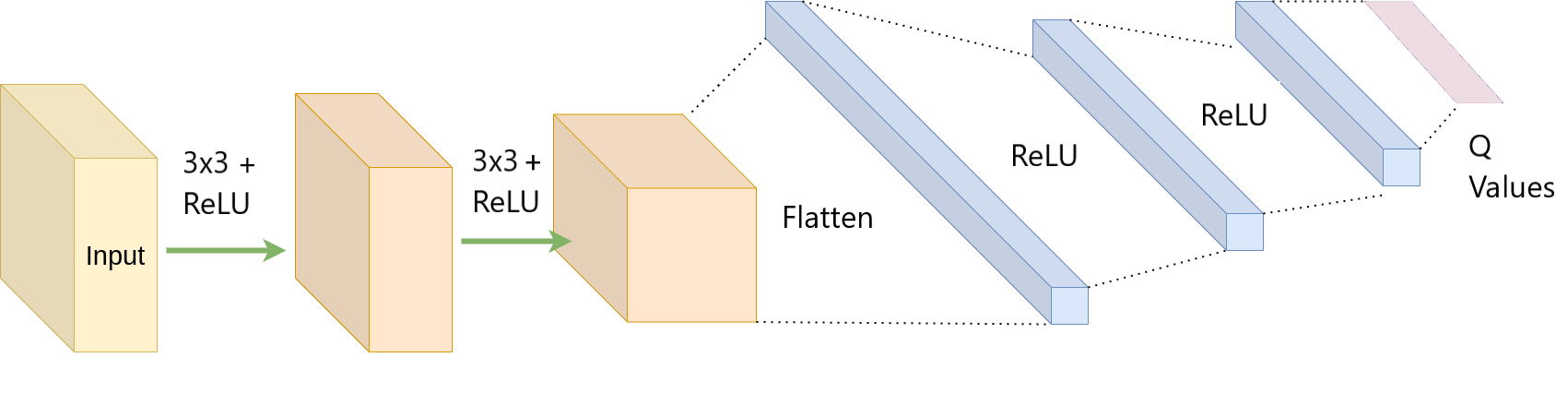}
    \caption{General architecture of DDQN}
    \label{fig:DDQN}
\end{figure}

We use the Centralized Training and Decentralised Execution (CTDE) \cite{lowe2017multi} framework for training the multi-agent DDQN. The stochastic policy for each drone $i$, $\phi^{d^{i}}_{\theta_{d}}$ : $\mathcal{O}^{i}_{d}$ x $A_{d}$ $\rightarrow$ [0, 1] is learnt using a DDQN, where $\theta_{d}$ represents the parameters of the DDQN shared by all drones. Similarly, a DDQN learns the stochastic policy for each ranger $j$, $\phi^{r^{j}}_{\theta_{r}}$ : $\mathcal{O}^{j}_{r}$ x $A_{r}$ $\rightarrow$ [0, 1] with parameters $\theta_{r}$ shared by all rangers. We thus use a centralized controller for all drones and a centralized controller for all rangers, sharing information with each other through the signaling actions of the drones. While each DDQN learns from the experiences of all the drones/rangers that it controls (centralized learning), actions for each drone/ranger are taken independently (decentralized execution).

The state representation used as input to the DDQN is a 3D Tensor with 9 channels, each with the same width and height as the grid: The current position of the drone/ranger in the grid, Whether an attacker has been detected in the current cell, Positions of all other drones, Positions of all other rangers, The outputs of the object detectors on each drone (0 or 1), Whether each drone is currently notifying rangers or not, Whether each drone is currently signaling or not, Animal density of the visited cell and  Visitation counts of each defender agent in the grid since the beginning of the episode. Fig. \ref{fig:DDQN} shows the architecture of the DDQN. More details about the neural network architecture are described in Supplementary Material. 
\subsection{Computing Allocation Strategy} 

We propose formulating the allocation stage as a two-player, zero-sum game. Computing a Stackelberg Security Equilibrium (SSE) strategy for the allocation stage would be the same as computing a NE strategy, since every NE is also a SSE in a two-player, zero-sum game \cite{fudenberg1998theory}. 

For a grid with $N$ cells, there are $\frac{N!}{(n_{d}+n_{r})!n_{d}!n_{r}!}$ possible pure strategy allocations for the defender. Computing a NE allocation strategy would involve searching in a huge, discrete and multi-dimensional space of all possible allocation strategies. For huge values of $N$, this computation quickly becomes intractable using traditional approaches based on LP and MILP. We therefore propose a gradient-based approximation approach using the policy gradient theorem and coPO.
 
However, the vanilla policy gradient algorithm performs poorly with large discrete action spaces. To address this, we can instead decompose a policy into a component that acts in a latent space of action representations(embeddings) and a component that transforms these representations into actual actions, as shown in \cite{chandak2019learning}. This allows generalization over actions, as similar actions have similar action representations, and improves performance, while speeding up learning. We therefore use learnt embeddings to represent allocations.

For a given environment and gridsize, we generate a dataset of 100000 possible allocations, which are possible actions for a player. We then train an autoencoder neural network on this dataset to learn latent space representations(embeddings) for these allocations. The network consists of two fully-connected dense layers with a \textit{tanh} activation after the first layer. Once training of the autoencoder is complete, we obtain embeddings $e^{d}$ and $e^{a}$  corresponding to defender and attacker allocations using the encoder part of this network.
 
We use neural networks to parameterize the attacker and defender allocation strategies, now given by $\pi^{d}(\textbf{w}^{d})$ and $\pi^{a}(\textbf{w}^{a}$), where $\textbf{w}^{d}$ and $\textbf{w}^{a}$ represent the weights of the neural networks. Details about the neural network architecture are provided in Supplementary Material. The defender and attacker allocation policy networks search over the space of embeddings $e^{d}$ and $e^{a}$, respectively. Once an embedding is sampled by either allocation policy network, it is mapped back to the allocation that it represents most closely and the patrolling stage is simulated to receive payoffs $R^{d}$ and $R^{a}$. The animal density distribution of the grid, represented by a 2D tensor and denoted by $s$, is fed as state information to these networks. The utilities of the defender and attacker ($U^{d}$ and $U^{a}$ = -$U^{d}$) are the expected rewards, given their policies:
\begin{equation}
\begin{split}
U^{d}(\textbf{w}^{d},\textbf{w}^{a}) = \mathbf{E}_{s,e^{d},e^{a}} [R^{d}(s,e^{d},e^{a})]\\
= \int_{s}\int_{e^{d}}\int_{e^{a}}\mathbf{P}(s)\pi^{d}(s|e^{d};\textbf{w}^{d})\pi^{a}(s|e^{a};\textbf{w}^{a})R^{d}(s,e^{d},e^{a})ds de^{d}de^{a}\\
\end{split}
\end{equation}

We then require to compute:

\begin{equation}
\textbf{w}^{d*} \epsilon arg \max_{\textbf{w}^{d}}\min_{\textbf{w}^{a}} U^{d}(\textbf{w}^{d},\textbf{w}^{a})
\end{equation}
\begin{equation}
\textbf{w}^{a*} \epsilon arg \min_{\textbf{w}^{a}}
U^{d}(\textbf{w}^{d*},\textbf{w}^{a})
\end{equation}

The weights are updated by coPO \cite{prajapat2020competitive} to arrive at an approximate NE for $\textbf{w}^{d*}$ and $\textbf{w}^{a*}$, as follows:
\begin{equation}
\begin{split}
\textbf{w}^{d} \leftarrow \textbf{w}^{d} +  \argmax_{\Delta\textbf{w}^{d}:\Delta\textbf{w}^{d}+\textbf{w}^{d}}{\Delta\textbf{w}^{d}}^{T} D_{\textbf{w}^{d}}U^{d} +\\{\Delta\textbf{w}^{d}}^{T} D_{\textbf{w}^{d}\textbf{w}^{a}}U^{d}\Delta\textbf{w}^{a} - \frac{1}{2\alpha}\parallel\Delta\textbf{w}^{d}\parallel^{2}
\end{split}
\end{equation}
\begin{equation}
\begin{split}
\textbf{w}^{a} \leftarrow \textbf{w}^{a} + \argmax_{\Delta\textbf{w}^{a}:\Delta\textbf{w}^{a}+\textbf{w}^{a}}{\Delta\textbf{w}^{a}}^{T} D_{\textbf{w}^{a}}U^{d} + \\{\Delta\textbf{w}^{a}}^{T} D_{\textbf{w}^{a}\textbf{w}^{d}}U^{d}\Delta\textbf{w}^{d} - \frac{1}{2\alpha}\parallel\Delta\textbf{w}^{a}\parallel^{2}
\end{split}
\end{equation}
where $\alpha$ represents the stepsize.

\subsection{CombSGPO Algorithm} 
We describe CombSGPO which combines computation of optimal strategies for allocation and patrolling. The training steps are summarized in Algorithm 1. First, we learn the defender patrolling strategy, choosing random allocations for the defender agents and the attackers at the start of each training episode and simulating the patrolling stage, to converge to a patrolling strategy $\phi^{d}$ for the defender agents against the attacker following $\phi^{a}$. Fig. 2 illustrates the CombSGPO pipeline.

For a given state $s$ of the park and pre-trained action embeddings $e^{d}$ and $e^{a}$, our algorithm solves the allocation stage as follows. At each episode, $n_{s}$ action embeddings are sampled from $\pi^{d}$ and $\pi^{a}$. Each action embedding is then matched to the closest allocation that it represents using a cosine similarity measure to get $n_s$ defender and attacker allocations. These allocations are then used as initial positions for the patrolling stage. The  simulator for the patrolling stage then runs $n_{s}$ patrols to completion, returning payoffs $R^{d}$ and $R^{a}$ for each pair of defender and attacker allocations. The sample allocations chosen and payoffs obtained are used to update $\pi^{d}$ and $\pi^{a}$ using coPO. We use $n_{s}$=10 in our experiments.

Self-play based approaches \cite{kamra2018policy, lanctot2017unified, heinrich2016deep} store the game state and sampled actions at each episode in a replay memory and then update each player's policy towards the best response to the other player's average policy. Our approach is more sample-efficient as it eliminates the need for a replay memory and directly updates each player's policy towards an approximate NE.

\begin{algorithm}[tp]

\SetAlgoLined
\SetKwInOut{Input}{Input}
\SetKwInOut{Output}{output}

\tcc{Train the DDQNs for drones and rangers to get patrolling strategies}
\Repeat{convergence}{
    Sample a random allocation for defenders and attackers\;
    Run the episode and sample experiences to DDQNs' buffers\;
    Update parameters $\theta_d$ and $\theta_r$ for DDQNs for defender patrolling strategy $\phi^{d}$\;
}

\tcc{Train defender and attacker allocation strategies}
Sample 100000 random allocations each for defenders and attackers in $\mathcal{D}_d$ and $\mathcal{D}_a$ respectively\;
Train autoencoders $f_d:\mathcal{D}_d \rightarrow \mathcal{R}^k$ and $f_a:\mathcal{D}_a \rightarrow \mathcal{R}^k$\;

Initialize $\mathbf{w}^d$ and $\mathbf{w}^a$\;
\Repeat{convergence}{
    Obtain allocation stage state = $s$\;
    \For{i = 1 to $n_s$}{
        Sample $e^d$ from $\pi^d(.|s_{i};\mathbf{w}^d)$ and $e^a$ from $\pi^a(.|s_{i};\mathbf{w}^a)$\;
        Get allocations $a^d_i=\underset{a\in \mathcal{D}_d}{\arg\min} ||f_d(a)-e^d||^2$ and $a^a_i=\underset{a\in \mathcal{D}_a}{\arg\min} ||f_a(a)-e^a||^2$\;
        Simulate patrolling stage with defender allocation $a^d_i$, attacker allocation $a^a_i$, defender patrolling strategy $\phi^{d}$, attacker patrolling strategy $\phi^{a}$ and recieve rewards $R^{d}_i$, $R^{a}_i$\;
    }
    Update parameters $\mathbf{w}^d$ and $\mathbf{w}^a$ by coPO;
}
\caption{CombSGPO}
\label{alg:combsgpo}
\end{algorithm}

%% file: experiments.tex
\begin{figure*}[tbp]
    \centering
    \begin{subfigure}{85mm}
      \centering
      \includegraphics[width=85mm]{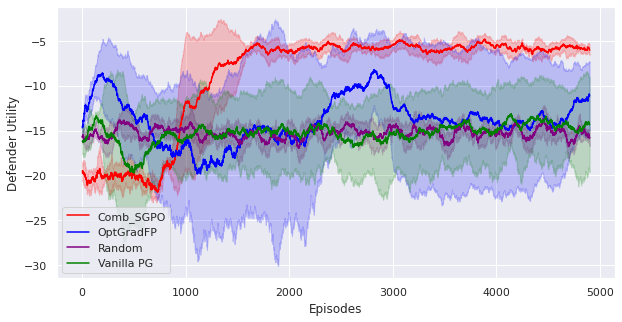}
    \end{subfigure}
    \begin{subfigure}{85mm}
      \centering
      \includegraphics[width=85mm]{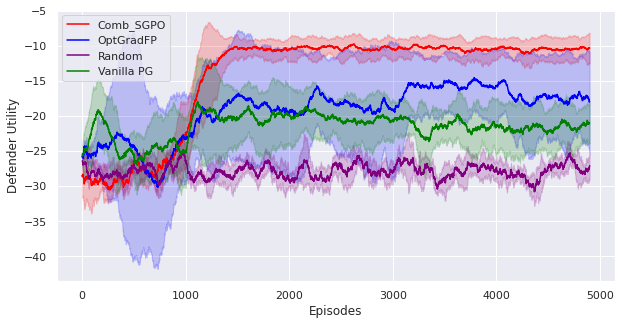}
    \end{subfigure}
   \caption{Training curves for experiments on a 15x15 grid ($\beta$=0,$\kappa$=0): SS (left) and MS (right)}
   \label{fig:results1}
\end{figure*}

\section{Experiments and Results}
We now present several experiments evaluating CombSGPO on the game model in Section 3. For the game, we present results for $n_d$ = 3 drones and $n_r$ = 2 rangers and gridsizes of 15x15 and 10x10. We ran experiments with $n_a$ = 1 as well as $n_a$ = 2, to show how the allocation and patrolling behavior changes when there is just one attacker compared to when there are multiple attackers, which requires splitting the team into sub-teams. We did this for random animal densities as well as spatial animal densities (i.e., based on distance to the border and other real-world park features like roads and rivers). For convenience, we denote the experiments by: SS (single attacker, spatial animal densities), SR (single attacker, random animal densities), MS (multiple attackers, spatial animal densities) and MR (multiple attackers, random animal densities).
Since we focused on real-world applicability, we focus primarily on SS and MS experiments in this section performed at different levels of uncertainty ($\beta$ and $\kappa$), while we defer SR and MR results to Supplementary Material. All results are averaged over five runs. Parameters used for training the neural networks are detailed in Supplementary Material.
We compared CombSGPO against the following baselines, all of which used a multi-agent DDQN for patrolling that was already trained by sampling random defender and attacker allocations at each episode:
 \begin{enumerate}
     \item Random: The defender allocations were assigned at random.
     \item Policy Gradient \cite{sutton2000policy}: We used this algorithm  to make players learn best responses to each other's strategies via policy gradient updates.
     \item OptGradFP \cite{kamra2018policy}: While this also uses policy gradient updates to learn best responses, the Fictitious Play \cite{lanctot2017unified} training framework is used to make both players converge to NE.
     \item GUARDSS \cite{bondi2020signal}: We replaced CombSGPO allocation with a GUARDSS allocation, then ran CombSGPO patrolling for a direct utility comparison. Specifically, we used the initial drone, ranger, and attacker locations (no signaling) from all pure strategies in the optimal mixed strategy for the GUARDSS model. GUARDSS only considers a single attacker, so MR and MS are left blank.
 \end{enumerate}

First, we evaluated the performance of CombSGPO against our baselines with different gridsizes, animal densities and number of attackers, with no uncertainty ($\beta$ = 0, $\kappa$ = 0). We sampled 150 episodes once the policies converged and averaged the defender utilities from these episodes. The training curves for SS and MS on a 15x15 grid are shown in Fig. \ref{fig:results1}. Average utilities for different experiments on 15x15 as well as 10x10 grids are shown in Tables \ref{tab:results1} and \ref{tab:results2}. CombSGPO significantly outperformed the baselines, converging faster and towards strategies with greater expected utility for the defender. We also observed that the variance of rewards in CombSGPO is much lower, indicating that robust strategies are learnt. 
\begin{table}[tp]
\centering
\begin{tabular}{|l|l|l|l|l|l|}
\hline
Setting & Random & PG     & OptGradFP & GUARDSS & CombSGPO            \\ \hline
SR      & -22.54 & -16.84 & -19.72    & -24.34                       & \textbf{-9.25} \\ \hline
MR     & -39.23 & -51.97 & -41.55   & -                      & \textbf{-24.49} \\ \hline
SS    & -15.35 & -14.99 & -14.49    & -14.11                          & \textbf{-5.53}  \\ \hline
MS     & -27.95 & -21.42 & -15.41    & -                          & \textbf{-10.39} \\ \hline
\end{tabular}
\caption{Average defender utilities for a 15x15 grid}
\label{tab:results1}
\begin{tabular}{|l|l|l|l|l|l|}
\hline
Setting & Random & PG     & OptGradFP & GUARDSS & CombSGPO            \\ \hline
SR      & -6.25 & -7.12 & -5.83    & -7.51                      & \textbf{-2.29} \\ \hline
MR     & -13.93 & -12.51 & -16.92    & -                       & \textbf{-7.72} \\ \hline
SS    & -5.37 & -4.55 & -3.47    & -6.03                          & \textbf{-1.63}  \\ \hline
MS     & -12.41 & -9.69 & -7.85    & -                          & \textbf{-4.65} \\ \hline
\end{tabular}
\caption{Average defender utilities for a 10x10 grid}
\label{tab:results2}
\end{table}

\begin{figure}[bp]
    \centering
    \includegraphics[width=80mm]{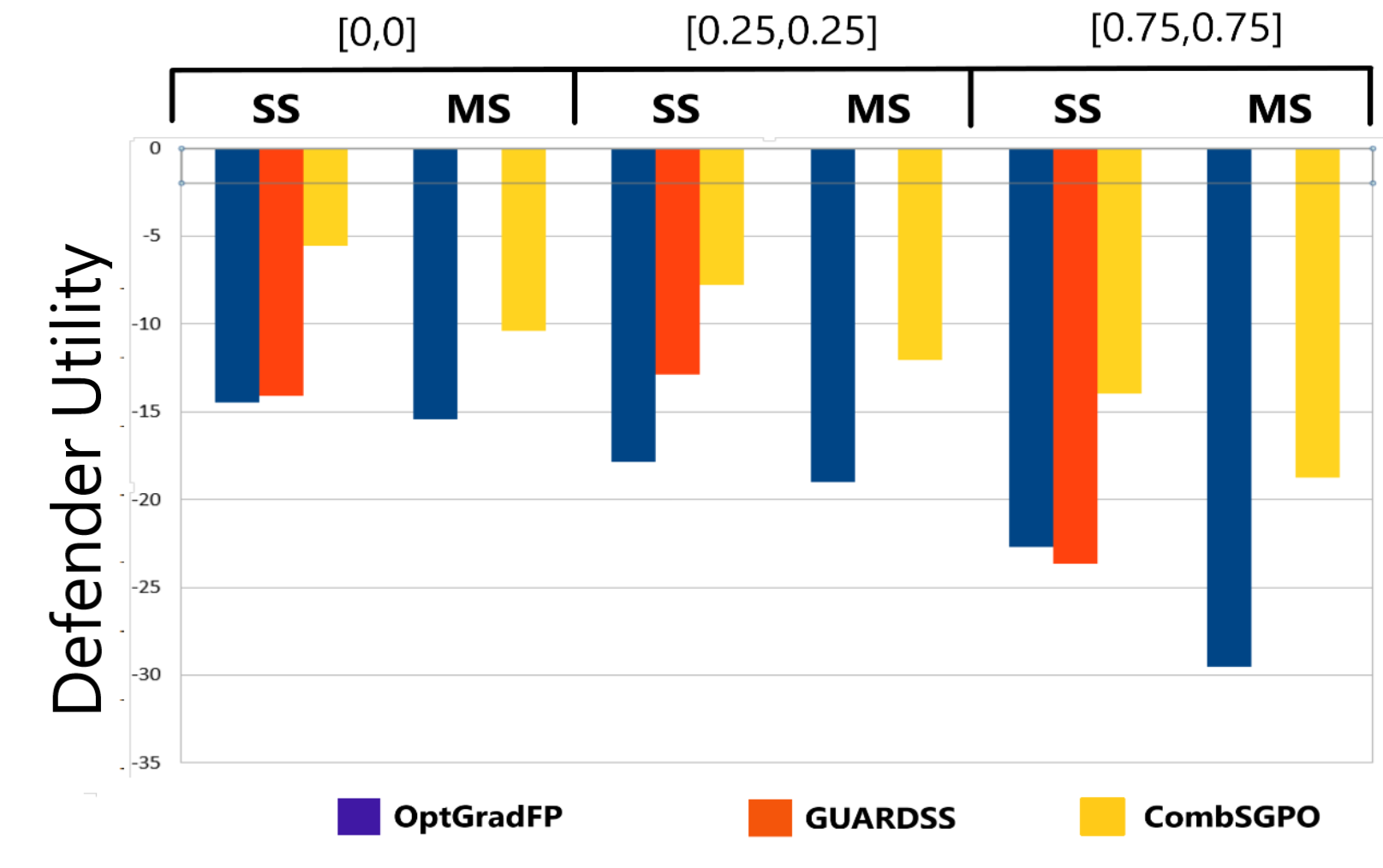}
    \caption{Average defender utilities under varying levels of [$\beta$, $\kappa$] }
    \label{fig:uc}
\end{figure}

Next, we evaluated how the perfomance of CombSGPO varied with uncertainty in the detectors and in the attackers' observations of drone signals. We performed experiments with three sets of values for $\beta$ and $\kappa$ : [0,0], [0.25,0.25] and [0.75,0.75], to simulate increasing levels of uncertainty. Defender utilities were calculated from 150 episodes sampled once the policies converged for SS and MS on a 15x15 grid. Only OptGradFP and GUARDSS have been considered, as the other baselines did not converge to an equilibrium. Results are shown in Fig. 5, with CombSGPO having the best performance, as it is most positive. In general, defender utility decreases with an increase in uncertainty.

For a 15x15 grid with SS, we show heatmaps of target cells attacked by sampling 100 attacker allocations and simulating the game, from the final strategies of the attacker for each algorithm, in Fig. 6. This is a direct measure of how effective the final learnt defender strategies of each algorithm are, as better defender strategies limit attacks to smaller regions of the grid and regions with lower animal densities. The animal density distribution is also shown alongside for comparison. 
In all algorithms considered, there was a clear tendency for the attacker to attack cells that were closer to the boundaries of the grid. While attackers in both OptGradFP and PG attacked multiple quadrants and larger regions overall, both CombSGPO and GUARDSS defenders were able to limit attacks mainly to one quadrant. Out of these two, CombSGPO defenders limited attacks to a smaller region.

\begin{figure}[tp]
    \centering
    \includegraphics[width=70mm]{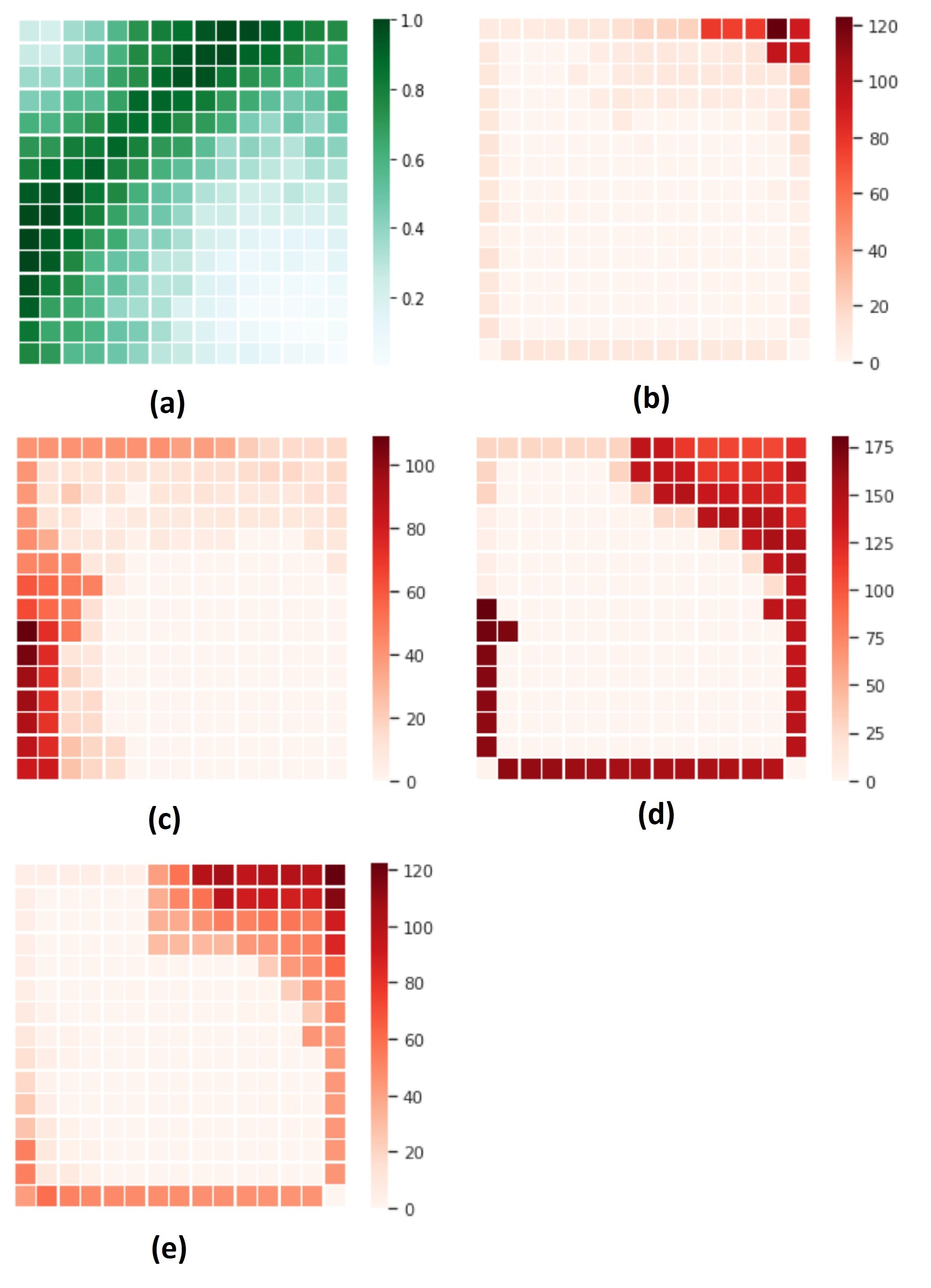}
    \caption{Visualization of sampled attacks (heatmaps). (a) shows the animal density distribution while the other figures show attacker heatmaps for: (b) CombSGPO, (c) GUARDSS, (d) OptGradFP and (e) PG. }
    \label{fig:uc}
\end{figure}

Having shown that CombSGPO outperformed all considered baselines quantitatively in terms of faster convergence, lesser variance, greater defender utilities and qualitatively in terms of greater protection to target cells, we  evaluated the time taken to converge to equilibrium for CombSGPO, OptGradFP and GUARDSS on a single core of Intel(R) Xeon(R) Platinum 8268 CPU @ 2.90GHz. Table 3 shows the results for SS on 15x15 and 10x10 grids, averaged over 5 runs, with CombSGPO performing best. 

\begin{table}[tp]
\centering
\begin{tabular}{|l|l|l|l|}
\hline
Gridsize & OptGradFP & GUARDSS & CombSGPO            \\ \hline
10x10    & 11.11$\pm$0.006  & 65.09$\pm$15.45 & \textbf{4.19$\pm$0.001}  \\ \hline
15x15     & 24.62$\pm$0.029 & 79.60$\pm$20.65 & \textbf{9.92$\pm$0.008}   \\ \hline
\end{tabular}
\caption{Timing results for 15x15 and 10x10 grids(in minutes)}
\label{tab:results3}
\end{table}

%% file: policies.tex
\section{Analysis of Policies Learnt By CombSGPO}

In this section, we provide a detailed qualitative analysis of the final strategies learnt by our trained CombSGPO model, including over multiple $\beta$ and $\kappa$. For better insight, we show snapshots taken from sampled episodes, illustrating the strategic behavior that we now describe. In all snapshots, green represents rangers, blue represents drones, red represents attackers, grey represents a signaling drone, and yellow represents a notifying drone. Both signaling and notifying actions are emphasized with a box around the snapshot. The animal density of each cell is shown by the shade of the colour in it (darker shades have higher animal density).

\begin{figure}[bp]
    \centering
    \includegraphics[width=\linewidth]{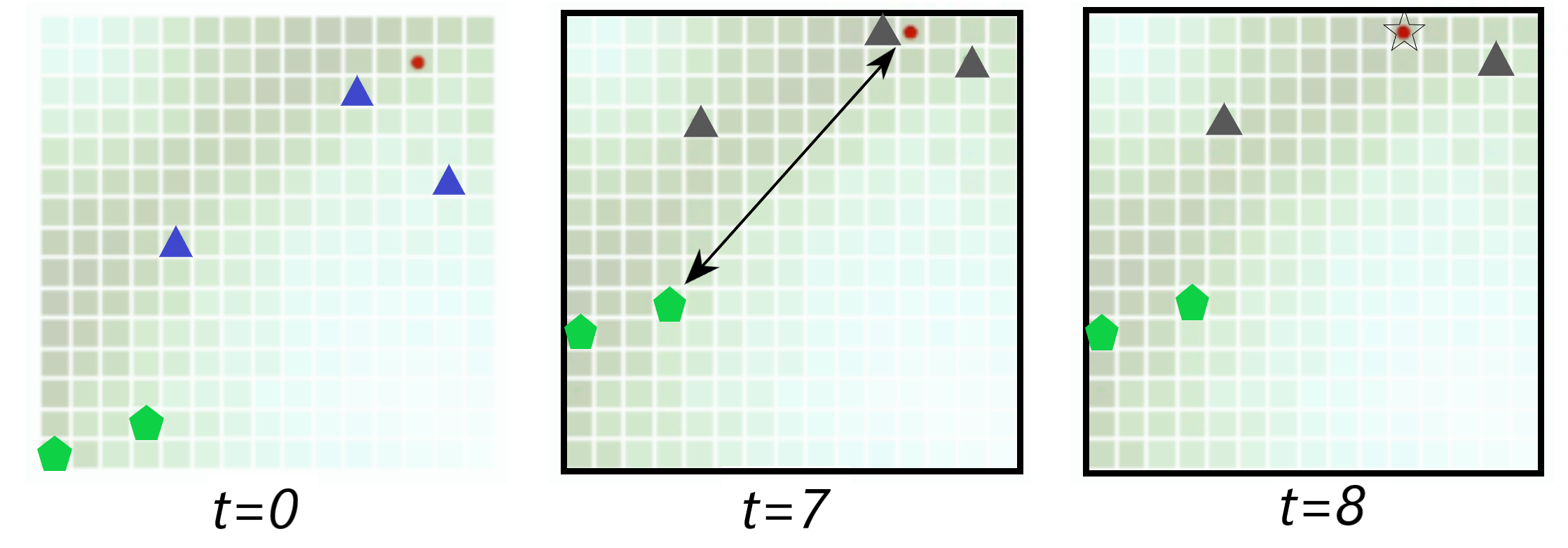}
    \caption{Sampled SS episode($\beta$=0,$\kappa$=0). See Section 6.}
\end{figure}

Throughout all experiments, we observed that CombSGPO learnt to deploy all drones together in one group and all rangers together in another group, as this allowed a greater area to be patrolled by the defender agents. We also observed that \emph{whenever a ranger was too far away from the drones to apprehend an attacker if the drone notified it, the drones learnt to signal the attacker and make it flee instead of notifying}, similar to the strategic signaling modeled by GUARDSS. This is illustrated in Fig. 7 which shows a sampled SS episode under $\beta$=0, $\kappa$=0. At t=0, we see that drones and rangers are deployed as separate groups, with the attacker deployed closer to the drones and the rangers are too far away from the attacker. At t=7, we can already see the drones moving closer to the attacker, while signaling at the same time. The arrow indicates that the rangers are far away from the attacker. The next frame at t=8 shows that the attacker is signalled by a drone in the same cell  (indicated by a star symbol), causing him to flee.

In some cases, we observed that defender agents allocated in a web-like manner around the attacker and closed in on the attacker, cornering him. The attractiveness of a cell to an attacker also depends on his distance from initial allocations of defender agents (as described in Section 4.1). Through this allocation pattern, \emph{the attacker was deliberately forced away from the areas of greater animal density and towards the corners and edges}. Fig 8. shows such a pattern for a sampled SR episode under $\beta$=0, $\kappa$=0 where defender agents deploy in a web-like formation (indicated by the black lines).

\begin{figure}[h]
    \centering
    \includegraphics[width=\linewidth]{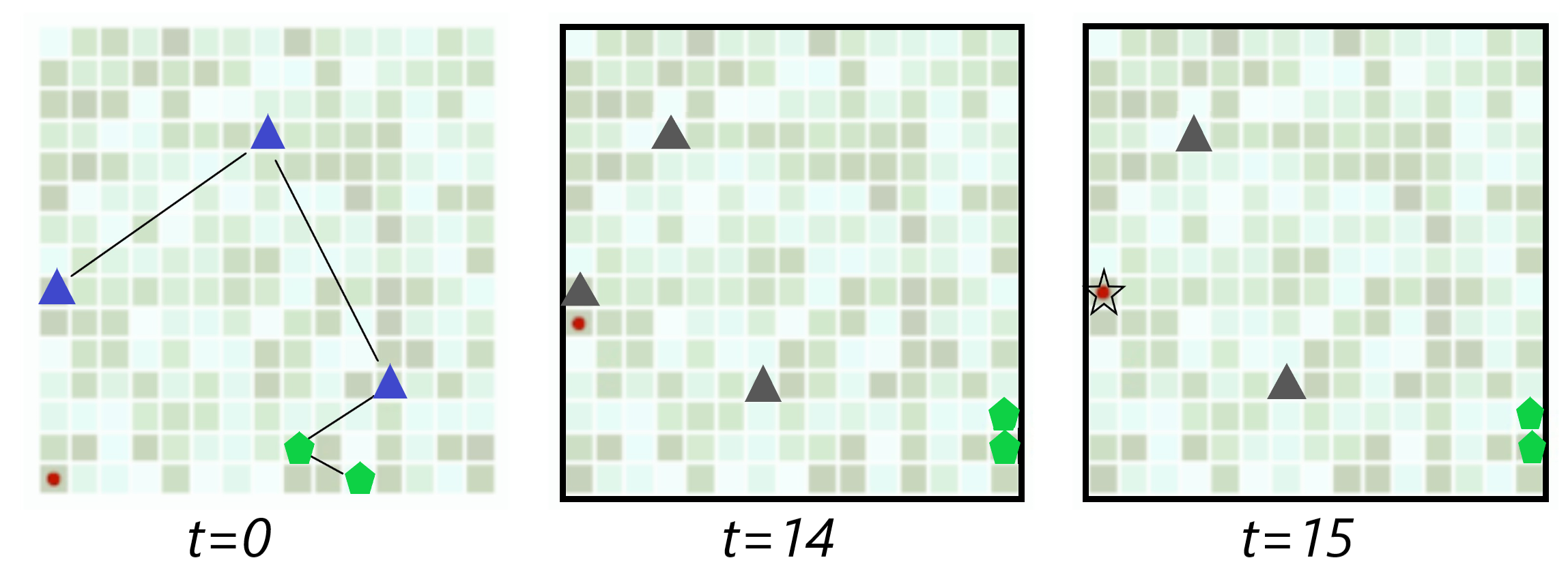}
    \caption{Sampled SR episode($\beta$=0,$\kappa$=0). See Section 6.}
\end{figure}
The allocation and signaling strategies were even clearer in the MR and MS experiments, as illustrated in Fig. 9, a sampled MS episode under $\beta$=0.25, $\kappa$=0.25. While one attacker is caught by the rangers (capture indicated by X), the other one, deployed in a completely different region, is forced to flee by the drones.
\begin{figure}[h]
    \centering
    \includegraphics[width=\linewidth]{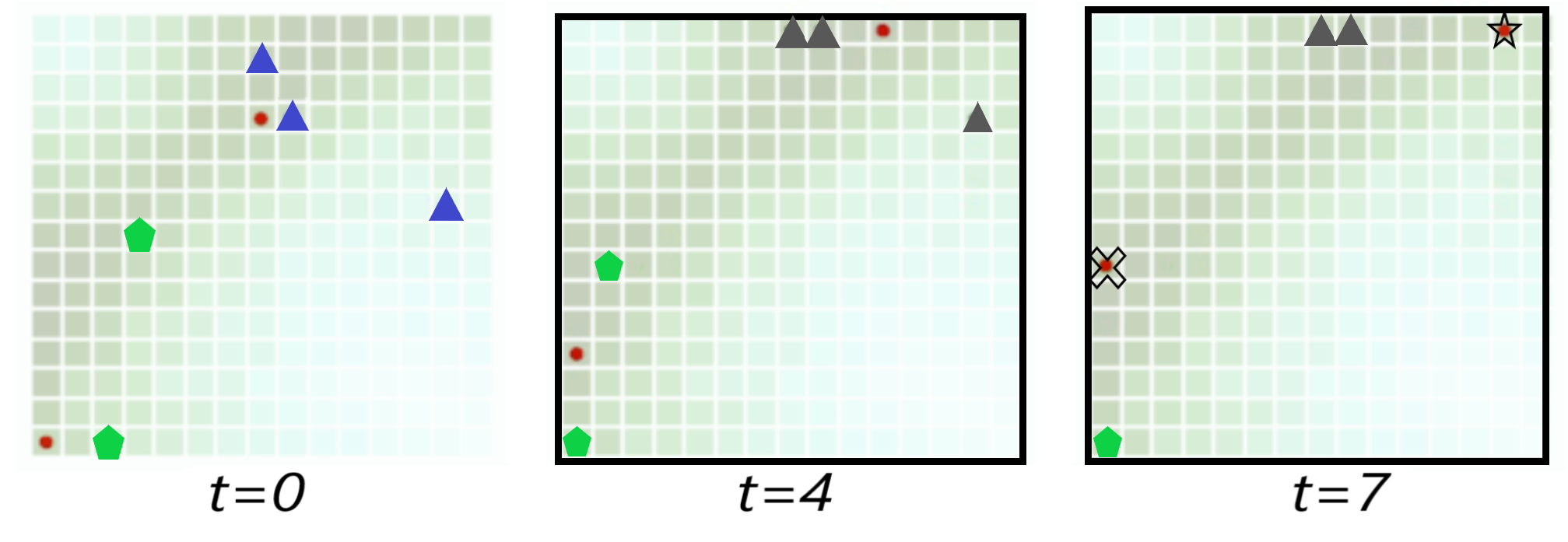}
    \caption{Sampled MS episode($\beta$=0.25,$\kappa$=0.25). See Section 6.}
\end{figure}

Apart from signaling attackers, notifications from drones to rangers were also observed. Interestingly, in a lot of cases where the attacker was allocated closer to the rangers than the drones, we noticed spurts of notification activity from the drones in the steps just before the attacker was caught, as illustrated in Fig. 10, a sampled SR episode under $\beta$=0.75, $\kappa$=0.75. This may indicate \emph{strategic notifications} to help guide the rangers in the correct direction. 

In addition to this, \emph{whenever the drones failed to detect the attacker even after many timesteps of patrolling, they responded with signaling and notifications}. In particular, they went into a state of constant signaling activity in the SS, MS and MR experiments and into a state of constant notification activity in SR experiments. Fig. 11 shows a sampled SS episode $\beta$=0.75, $\kappa$=0.75 where the drones signal after 13 timesteps. This is similar in spirit to the ranger moving in the GUARDSS model when an attacker is not seen for some time, called matching. In our model, this could indicate that drones are trying to compensate for possible false negative detections, or possibly that drones are suggesting that attackers are near rangers instead. 

\begin{figure}[h]
    \centering
    \includegraphics[width=\linewidth]{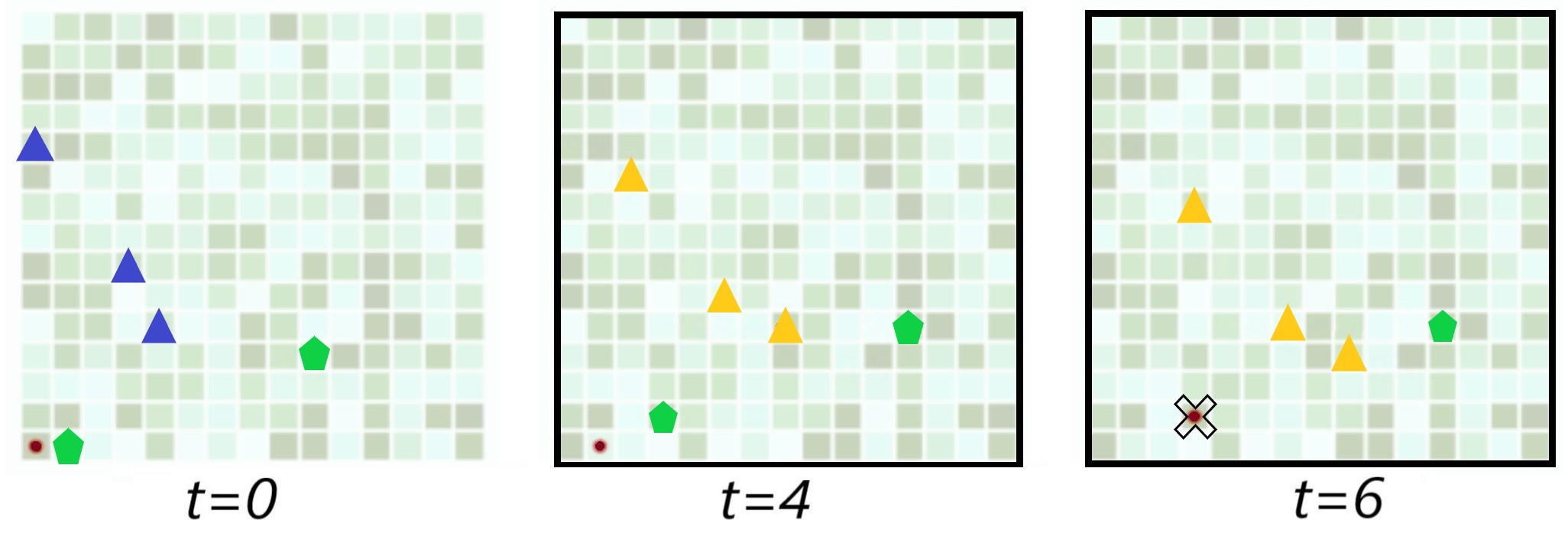}
    \caption{Sampled SS episode($\beta$=0.75,$\kappa$=0.75). See Section 6.}
\end{figure}

\begin{figure}[h]
    \centering
    \includegraphics[width=\linewidth]{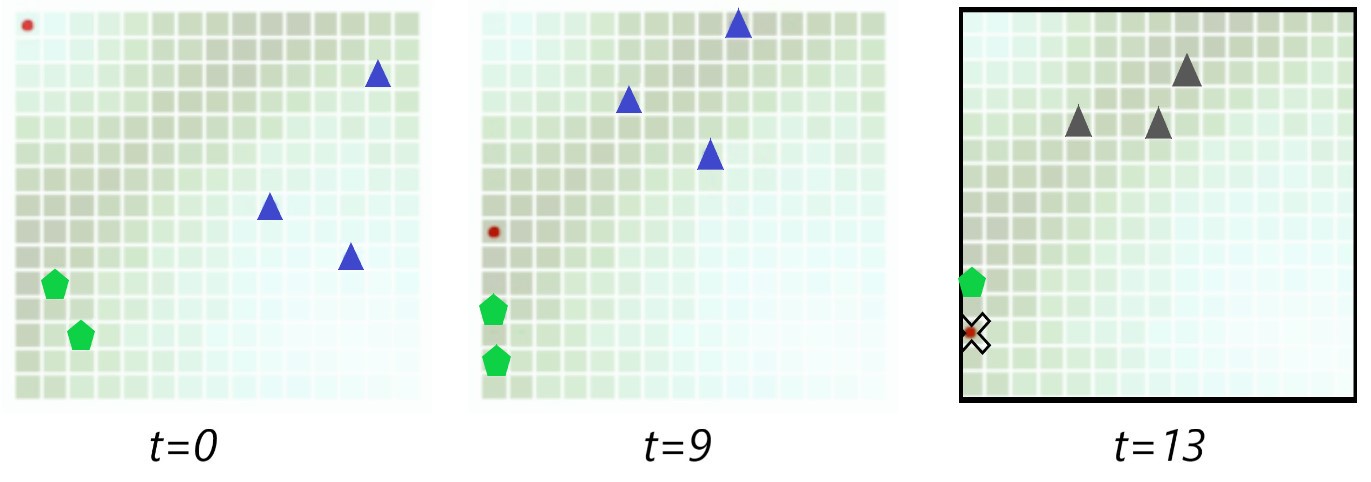}
    \caption{Sampled SS episode($\beta$=0.25,$\kappa$=0.25). See Section 6.}
\end{figure}

As uncertainty increases from $\beta$=0, $\kappa$=0 to $\beta$=0.75, $\kappa$=0.75, we see a marked decrease in the amount of signaling and notification activity in SR experiments while the amount of signaling and notification activity stays the same in SS, MS and MR experiments.


%
%


%% file: discussions.tex
\section{Related Work}
In RL for GSGs, \cite{kamra2018policy,kamra2019deepfp} use Policy Gradient learning for resource allocation and \cite{wang2019deep} uses an algorithm based on Deep Q-Learning \cite{mnih2013playing} for computing patrolling strategies with fixed initial allocations. While \cite{wang2019deep} does model real-time information, it does not incorporate uncertainty or strategic signalling with multiple defender resources and attackers into this model.

Bondi et al. \cite{bondi2020signal} also model uncertainty in real-time information in an application that uses conservation drones to prevent illegal wildlife poaching. Drones are able to provide real-time, uncertain information to defenders, as drones are fitted with thermal infrared cameras, and an object detection system called Systematic POacher deTector (SPOT) \cite{bondi2018spot} provides (uncertain) automatic detection of potential poachers (attackers) and animals. If suspicious activity is detected, nearby park rangers (defenders) can be mobilized to respond to the situation. Warning signals may also be deployed by the drones to deter poachers, which consists of turning the lights on aboard the drone to signal to poachers that rangers are on their way. Signaling can be done deceptively in order to lead poachers to believe that park rangers may respond, even when rangers are unable to respond, for example due to distance. It is to be noted that while this approach models allocation and signalling decisions, it does not model patrolling strategies. Additionally, it is not capable of modelling multiple attackers and is not efficiently scalable to larger parks (areas to protect). 

In contrast, our work proposes a unified solution to resource allocation and patrolling. We also focus on strategic signaling and communication between agents for coordinated patrolling in the presence of uncertain, real-time information, which is an area that is yet to be addressed properly in complex GSGs by using RL methods.
\section{Discussion and Conclusion}

We introduce a novel solution approach based on reinforcement learning for computing defender strategies in a two-stage GSG, that combines resource allocation, patrolling, communication between resources and signaling in the presence of uncertain real-time information. We showed, both quantitatively and qualitatively, that our approach outperforms comparable approaches over different types of environments that simulate green security domains, including protecting animals from illegal wildlife poaching. We also showed that our trained models learn strategic behavior such as formation of sub-teams within a team of drones and rangers, coordinated patrolling formations and strategic and even deceptive signaling in the presence of uncertainty to ward off and/or apprehend poachers.

An interesting future direction would be the replacement of the binary object detectors on drones with real-time image input from object detectors on drone cameras. This would introduce intrinsic uncertainty into the model and also take us a step closer to deploying such an application in real-world scenarios.

%% file: sample.bbl

\begin{thebibliography}{26}


\ifx \showCODEN    \undefined \def \showCODEN     #1{\unskip}     \fi
\ifx \showDOI      \undefined \def \showDOI       #1{#1}\fi
\ifx \showISBNx    \undefined \def \showISBNx     #1{\unskip}     \fi
\ifx \showISBNxiii \undefined \def \showISBNxiii  #1{\unskip}     \fi
\ifx \showISSN     \undefined \def \showISSN      #1{\unskip}     \fi
\ifx \showLCCN     \undefined \def \showLCCN      #1{\unskip}     \fi
\ifx \shownote     \undefined \def \shownote      #1{#1}          \fi
\ifx \showarticletitle \undefined \def \showarticletitle #1{#1}   \fi
\ifx \showURL      \undefined \def \showURL       {\relax}        \fi
\providecommand\bibfield[2]{#2}
\providecommand\bibinfo[2]{#2}
\providecommand\natexlab[1]{#1}
\providecommand\showeprint[2][]{arXiv:#2}

\bibitem[\protect\citeauthoryear{Basak, Fang, Nguyen, and Kiekintveld}{Basak
  et~al\mbox{.}}{2016}]%
        {basak2016combining}
\bibfield{author}{\bibinfo{person}{Anjon Basak}, \bibinfo{person}{Fei Fang},
  \bibinfo{person}{Thanh~Hong Nguyen}, {and} \bibinfo{person}{Christopher
  Kiekintveld}.} \bibinfo{year}{2016}\natexlab{}.
\newblock \showarticletitle{Combining graph contraction and strategy generation
  for green security games}. In \bibinfo{booktitle}{\emph{International
  Conference on Decision and Game Theory for Security}}. Springer,
  \bibinfo{pages}{251--271}.
\newblock


\bibitem[\protect\citeauthoryear{Basilico, De~Nittis, and Gatti}{Basilico
  et~al\mbox{.}}{2015}]%
        {basilico2015security}
\bibfield{author}{\bibinfo{person}{Nicola Basilico}, \bibinfo{person}{Giuseppe
  De~Nittis}, {and} \bibinfo{person}{Nicola Gatti}.}
  \bibinfo{year}{2015}\natexlab{}.
\newblock \showarticletitle{A security game model for environment protection in
  the presence of an alarm system}. In \bibinfo{booktitle}{\emph{International
  Conference on Decision and Game Theory for Security}}. Springer,
  \bibinfo{pages}{192--207}.
\newblock


\bibitem[\protect\citeauthoryear{Bondi, Fang, Hamilton, Kar, Dmello, Choi,
  Hannaford, Iyer, Joppa, Tambe, et~al\mbox{.}}{Bondi et~al\mbox{.}}{2018}]%
        {bondi2018spot}
\bibfield{author}{\bibinfo{person}{Elizabeth Bondi}, \bibinfo{person}{Fei
  Fang}, \bibinfo{person}{Mark Hamilton}, \bibinfo{person}{Debarun Kar},
  \bibinfo{person}{Donnabell Dmello}, \bibinfo{person}{Jongmoo Choi},
  \bibinfo{person}{Robert Hannaford}, \bibinfo{person}{Arvind Iyer},
  \bibinfo{person}{Lucas Joppa}, \bibinfo{person}{Milind Tambe},
  {et~al\mbox{.}}} \bibinfo{year}{2018}\natexlab{}.
\newblock \showarticletitle{Spot poachers in action: Augmenting conservation
  drones with automatic detection in near real time.}. In
  \bibinfo{booktitle}{\emph{IAAI}}. \bibinfo{pages}{7741--7746}.
\newblock


\bibitem[\protect\citeauthoryear{Bondi, Oh, Xu, Fang, Dilkina, and Tambe}{Bondi
  et~al\mbox{.}}{2020}]%
        {bondi2020signal}
\bibfield{author}{\bibinfo{person}{Elizabeth Bondi}, \bibinfo{person}{Hoon Oh},
  \bibinfo{person}{Haifeng Xu}, \bibinfo{person}{Fei Fang},
  \bibinfo{person}{Bistra Dilkina}, {and} \bibinfo{person}{Milind Tambe}.}
  \bibinfo{year}{2020}\natexlab{}.
\newblock \showarticletitle{To Signal or Not To Signal: Exploiting Uncertain
  Real-Time Information in Signaling Games for Security and Sustainability.}.
  In \bibinfo{booktitle}{\emph{AAAI}}. \bibinfo{pages}{1369--1377}.
\newblock


\bibitem[\protect\citeauthoryear{Chandak, Theocharous, Kostas, Jordan, and
  Thomas}{Chandak et~al\mbox{.}}{2019}]%
        {chandak2019learning}
\bibfield{author}{\bibinfo{person}{Yash Chandak}, \bibinfo{person}{Georgios
  Theocharous}, \bibinfo{person}{James Kostas}, \bibinfo{person}{Scott Jordan},
  {and} \bibinfo{person}{Philip~S Thomas}.} \bibinfo{year}{2019}\natexlab{}.
\newblock \showarticletitle{Learning action representations for reinforcement
  learning}.
\newblock \bibinfo{journal}{\emph{arXiv preprint arXiv:1902.00183}}
  (\bibinfo{year}{2019}).
\newblock


\bibitem[\protect\citeauthoryear{Fang, Nguyen, Pickles, Lam, Clements, An,
  Singh, Tambe, Lemieux, et~al\mbox{.}}{Fang et~al\mbox{.}}{2016}]%
        {fang2016deploying}
\bibfield{author}{\bibinfo{person}{Fei Fang}, \bibinfo{person}{Thanh~Hong
  Nguyen}, \bibinfo{person}{Rob Pickles}, \bibinfo{person}{Wai~Y Lam},
  \bibinfo{person}{Gopalasamy~R Clements}, \bibinfo{person}{Bo An},
  \bibinfo{person}{Amandeep Singh}, \bibinfo{person}{Milind Tambe},
  \bibinfo{person}{Andrew Lemieux}, {et~al\mbox{.}}}
  \bibinfo{year}{2016}\natexlab{}.
\newblock \showarticletitle{Deploying PAWS: Field Optimization of the
  Protection Assistant for Wildlife Security.}. In
  \bibinfo{booktitle}{\emph{AAAI}}, Vol.~\bibinfo{volume}{16}.
  \bibinfo{pages}{3966--3973}.
\newblock


\bibitem[\protect\citeauthoryear{Fang, Stone, and Tambe}{Fang
  et~al\mbox{.}}{2015}]%
        {fang2015security}
\bibfield{author}{\bibinfo{person}{Fei Fang}, \bibinfo{person}{Peter Stone},
  {and} \bibinfo{person}{Milind Tambe}.} \bibinfo{year}{2015}\natexlab{}.
\newblock \showarticletitle{When Security Games Go Green: Designing Defender
  Strategies to Prevent Poaching and Illegal Fishing.}. In
  \bibinfo{booktitle}{\emph{IJCAI}}. \bibinfo{pages}{2589--2595}.
\newblock


\bibitem[\protect\citeauthoryear{Fudenberg, Drew, Levine, and Levine}{Fudenberg
  et~al\mbox{.}}{1998}]%
        {fudenberg1998theory}
\bibfield{author}{\bibinfo{person}{Drew Fudenberg}, \bibinfo{person}{Fudenberg
  Drew}, \bibinfo{person}{David~K Levine}, {and} \bibinfo{person}{David~K
  Levine}.} \bibinfo{year}{1998}\natexlab{}.
\newblock \bibinfo{booktitle}{\emph{The theory of learning in games}}.
  Vol.~\bibinfo{volume}{2}.
\newblock \bibinfo{publisher}{MIT press}.
\newblock


\bibitem[\protect\citeauthoryear{Heinrich and Silver}{Heinrich and
  Silver}{2016}]%
        {heinrich2016deep}
\bibfield{author}{\bibinfo{person}{Johannes Heinrich} {and}
  \bibinfo{person}{David Silver}.} \bibinfo{year}{2016}\natexlab{}.
\newblock \showarticletitle{Deep reinforcement learning from self-play in
  imperfect-information games}.
\newblock \bibinfo{journal}{\emph{arXiv preprint arXiv:1603.01121}}
  (\bibinfo{year}{2016}).
\newblock


\bibitem[\protect\citeauthoryear{Kamra, Gupta, Fang, Liu, and Tambe}{Kamra
  et~al\mbox{.}}{2018}]%
        {kamra2018policy}
\bibfield{author}{\bibinfo{person}{Nitin Kamra}, \bibinfo{person}{Umang Gupta},
  \bibinfo{person}{Fei Fang}, \bibinfo{person}{Yan Liu}, {and}
  \bibinfo{person}{Milind Tambe}.} \bibinfo{year}{2018}\natexlab{}.
\newblock \showarticletitle{Policy Learning for Continuous Space Security Games
  Using Neural Networks.}. In \bibinfo{booktitle}{\emph{AAAI}}.
  \bibinfo{pages}{1103--1112}.
\newblock


\bibitem[\protect\citeauthoryear{Kamra, Gupta, Wang, Fang, Liu, and
  Tambe}{Kamra et~al\mbox{.}}{2019}]%
        {kamra2019deepfp}
\bibfield{author}{\bibinfo{person}{Nitin Kamra}, \bibinfo{person}{Umang Gupta},
  \bibinfo{person}{Kai Wang}, \bibinfo{person}{Fei Fang}, \bibinfo{person}{Yan
  Liu}, {and} \bibinfo{person}{Milind Tambe}.} \bibinfo{year}{2019}\natexlab{}.
\newblock \showarticletitle{DeepFP for Finding Nash Equilibrium in Continuous
  Action Spaces}. In \bibinfo{booktitle}{\emph{International Conference on
  Decision and Game Theory for Security}}. Springer, \bibinfo{pages}{238--258}.
\newblock


\bibitem[\protect\citeauthoryear{Korzhyk, Yin, Kiekintveld, Conitzer, and
  Tambe}{Korzhyk et~al\mbox{.}}{2011}]%
        {korzhyk2011stackelberg}
\bibfield{author}{\bibinfo{person}{Dmytro Korzhyk}, \bibinfo{person}{Zhengyu
  Yin}, \bibinfo{person}{Christopher Kiekintveld}, \bibinfo{person}{Vincent
  Conitzer}, {and} \bibinfo{person}{Milind Tambe}.}
  \bibinfo{year}{2011}\natexlab{}.
\newblock \showarticletitle{Stackelberg vs. Nash in security games: An extended
  investigation of interchangeability, equivalence, and uniqueness}.
\newblock \bibinfo{journal}{\emph{Journal of Artificial Intelligence Research}}
   \bibinfo{volume}{41} (\bibinfo{year}{2011}), \bibinfo{pages}{297--327}.
\newblock


\bibitem[\protect\citeauthoryear{Kraus}{Kraus}{[n.d.]}]%
        {krausarmor}
\bibfield{author}{\bibinfo{person}{Sarit Kraus}.}
  \bibinfo{year}{[n.d.]}\natexlab{}.
\newblock \showarticletitle{ARMOR Software: A Game Theoretic Approach for
  Airport Security James Pita, Manish Jain, Fernando Ord{\'o}{\~n}ez,
  Christopher Portway, Milind Tambe, Craig Western University of Southern
  California, Los Angeles, CA 90089 Praveen Paruchuri Intelligent Automation,
  Inc., Rockville, MD 20855}.
\newblock  (\bibinfo{year}{[n.\,d.]}).
\newblock


\bibitem[\protect\citeauthoryear{Lanctot, Zambaldi, Gruslys, Lazaridou, Tuyls,
  P{\'e}rolat, Silver, and Graepel}{Lanctot et~al\mbox{.}}{2017}]%
        {lanctot2017unified}
\bibfield{author}{\bibinfo{person}{Marc Lanctot}, \bibinfo{person}{Vinicius
  Zambaldi}, \bibinfo{person}{Audrunas Gruslys}, \bibinfo{person}{Angeliki
  Lazaridou}, \bibinfo{person}{Karl Tuyls}, \bibinfo{person}{Julien
  P{\'e}rolat}, \bibinfo{person}{David Silver}, {and} \bibinfo{person}{Thore
  Graepel}.} \bibinfo{year}{2017}\natexlab{}.
\newblock \showarticletitle{A unified game-theoretic approach to multiagent
  reinforcement learning}. In \bibinfo{booktitle}{\emph{Advances in neural
  information processing systems}}. \bibinfo{pages}{4190--4203}.
\newblock


\bibitem[\protect\citeauthoryear{Lowe, Wu, Tamar, Harb, Abbeel, and
  Mordatch}{Lowe et~al\mbox{.}}{2017}]%
        {lowe2017multi}
\bibfield{author}{\bibinfo{person}{Ryan Lowe}, \bibinfo{person}{Yi~I Wu},
  \bibinfo{person}{Aviv Tamar}, \bibinfo{person}{Jean Harb},
  \bibinfo{person}{OpenAI~Pieter Abbeel}, {and} \bibinfo{person}{Igor
  Mordatch}.} \bibinfo{year}{2017}\natexlab{}.
\newblock \showarticletitle{Multi-agent actor-critic for mixed
  cooperative-competitive environments}. In \bibinfo{booktitle}{\emph{Advances
  in neural information processing systems}}. \bibinfo{pages}{6379--6390}.
\newblock


\bibitem[\protect\citeauthoryear{Mnih, Kavukcuoglu, Silver, Graves, Antonoglou,
  Wierstra, and Riedmiller}{Mnih et~al\mbox{.}}{2013}]%
        {mnih2013playing}
\bibfield{author}{\bibinfo{person}{Volodymyr Mnih}, \bibinfo{person}{Koray
  Kavukcuoglu}, \bibinfo{person}{David Silver}, \bibinfo{person}{Alex Graves},
  \bibinfo{person}{Ioannis Antonoglou}, \bibinfo{person}{Daan Wierstra}, {and}
  \bibinfo{person}{Martin Riedmiller}.} \bibinfo{year}{2013}\natexlab{}.
\newblock \showarticletitle{Playing atari with deep reinforcement learning}.
\newblock \bibinfo{journal}{\emph{arXiv preprint arXiv:1312.5602}}
  (\bibinfo{year}{2013}).
\newblock


\bibitem[\protect\citeauthoryear{Pita, Jain, Marecki, Ord{\'o}{\~n}ez, Portway,
  Tambe, Western, Paruchuri, and Kraus}{Pita et~al\mbox{.}}{2008}]%
        {pita2008deployed}
\bibfield{author}{\bibinfo{person}{James Pita}, \bibinfo{person}{Manish Jain},
  \bibinfo{person}{Janusz Marecki}, \bibinfo{person}{Fernando Ord{\'o}{\~n}ez},
  \bibinfo{person}{Christopher Portway}, \bibinfo{person}{Milind Tambe},
  \bibinfo{person}{Craig Western}, \bibinfo{person}{Praveen Paruchuri}, {and}
  \bibinfo{person}{Sarit Kraus}.} \bibinfo{year}{2008}\natexlab{}.
\newblock \showarticletitle{Deployed ARMOR protection: the application of a
  game theoretic model for security at the Los Angeles International Airport}.
  In \bibinfo{booktitle}{\emph{Proceedings of the 7th international joint
  conference on Autonomous agents and multiagent systems: industrial track}}.
  \bibinfo{pages}{125--132}.
\newblock


\bibitem[\protect\citeauthoryear{Prajapat, Azizzadenesheli, Liniger, Yue, and
  Anandkumar}{Prajapat et~al\mbox{.}}{2020}]%
        {prajapat2020competitive}
\bibfield{author}{\bibinfo{person}{Manish Prajapat}, \bibinfo{person}{Kamyar
  Azizzadenesheli}, \bibinfo{person}{Alexander Liniger},
  \bibinfo{person}{Yisong Yue}, {and} \bibinfo{person}{Anima Anandkumar}.}
  \bibinfo{year}{2020}\natexlab{}.
\newblock \showarticletitle{Competitive Policy Optimization}.
\newblock \bibinfo{journal}{\emph{arXiv preprint arXiv:2006.10611}}
  (\bibinfo{year}{2020}).
\newblock


\bibitem[\protect\citeauthoryear{Sutton, McAllester, Singh, and Mansour}{Sutton
  et~al\mbox{.}}{2000}]%
        {sutton2000policy}
\bibfield{author}{\bibinfo{person}{Richard~S Sutton}, \bibinfo{person}{David~A
  McAllester}, \bibinfo{person}{Satinder~P Singh}, {and}
  \bibinfo{person}{Yishay Mansour}.} \bibinfo{year}{2000}\natexlab{}.
\newblock \showarticletitle{Policy gradient methods for reinforcement learning
  with function approximation}. In \bibinfo{booktitle}{\emph{Advances in neural
  information processing systems}}. \bibinfo{pages}{1057--1063}.
\newblock


\bibitem[\protect\citeauthoryear{Tambe}{Tambe}{2011}]%
        {tambe2011security}
\bibfield{author}{\bibinfo{person}{Milind Tambe}.}
  \bibinfo{year}{2011}\natexlab{}.
\newblock \bibinfo{booktitle}{\emph{Security and game theory: algorithms,
  deployed systems, lessons learned}}.
\newblock \bibinfo{publisher}{Cambridge university press}.
\newblock


\bibitem[\protect\citeauthoryear{Tsai, Rathi, Kiekintveld, Ordonez, and
  Tambe}{Tsai et~al\mbox{.}}{2009}]%
        {tsai2009iris}
\bibfield{author}{\bibinfo{person}{Jason Tsai}, \bibinfo{person}{Shyamsunder
  Rathi}, \bibinfo{person}{Christopher Kiekintveld}, \bibinfo{person}{Fernando
  Ordonez}, {and} \bibinfo{person}{Milind Tambe}.}
  \bibinfo{year}{2009}\natexlab{}.
\newblock \showarticletitle{IRIS-a tool for strategic security allocation in
  transportation networks}.
\newblock \bibinfo{journal}{\emph{AAMAS (Industry Track)}}
  (\bibinfo{year}{2009}), \bibinfo{pages}{37--44}.
\newblock


\bibitem[\protect\citeauthoryear{Van~Hasselt, Guez, and Silver}{Van~Hasselt
  et~al\mbox{.}}{2015}]%
        {van2015deep}
\bibfield{author}{\bibinfo{person}{Hado Van~Hasselt}, \bibinfo{person}{Arthur
  Guez}, {and} \bibinfo{person}{David Silver}.}
  \bibinfo{year}{2015}\natexlab{}.
\newblock \showarticletitle{Deep reinforcement learning with double
  q-learning}.
\newblock \bibinfo{journal}{\emph{arXiv preprint arXiv:1509.06461}}
  (\bibinfo{year}{2015}).
\newblock


\bibitem[\protect\citeauthoryear{Wang, Shi, Yu, Wu, Singh, Joppa, and
  Fang}{Wang et~al\mbox{.}}{2019}]%
        {wang2019deep}
\bibfield{author}{\bibinfo{person}{Yufei Wang}, \bibinfo{person}{Zheyuan~Ryan
  Shi}, \bibinfo{person}{Lantao Yu}, \bibinfo{person}{Yi Wu},
  \bibinfo{person}{Rohit Singh}, \bibinfo{person}{Lucas Joppa}, {and}
  \bibinfo{person}{Fei Fang}.} \bibinfo{year}{2019}\natexlab{}.
\newblock \showarticletitle{Deep reinforcement learning for green security
  games with real-time information}. In \bibinfo{booktitle}{\emph{Proceedings
  of the AAAI Conference on Artificial Intelligence}},
  Vol.~\bibinfo{volume}{33}. \bibinfo{pages}{1401--1408}.
\newblock


\bibitem[\protect\citeauthoryear{Xu, Ford, Fang, Dilkina, Plumptre, Tambe,
  Driciru, Wanyama, Rwetsiba, Nsubaga, et~al\mbox{.}}{Xu et~al\mbox{.}}{2017}]%
        {xu2017optimal}
\bibfield{author}{\bibinfo{person}{Haifeng Xu}, \bibinfo{person}{Benjamin
  Ford}, \bibinfo{person}{Fei Fang}, \bibinfo{person}{Bistra Dilkina},
  \bibinfo{person}{Andrew Plumptre}, \bibinfo{person}{Milind Tambe},
  \bibinfo{person}{Margaret Driciru}, \bibinfo{person}{Fred Wanyama},
  \bibinfo{person}{Aggrey Rwetsiba}, \bibinfo{person}{Mustapha Nsubaga},
  {et~al\mbox{.}}} \bibinfo{year}{2017}\natexlab{}.
\newblock \showarticletitle{Optimal patrol planning for green security games
  with black-box attackers}. In \bibinfo{booktitle}{\emph{International
  Conference on Decision and Game Theory for Security}}. Springer,
  \bibinfo{pages}{458--477}.
\newblock


\bibitem[\protect\citeauthoryear{Xu, Wang, Vayanos, and Tambe}{Xu
  et~al\mbox{.}}{2018}]%
        {xu2018strategic}
\bibfield{author}{\bibinfo{person}{Haifeng Xu}, \bibinfo{person}{Kai Wang},
  \bibinfo{person}{Phebe Vayanos}, {and} \bibinfo{person}{Milind Tambe}.}
  \bibinfo{year}{2018}\natexlab{}.
\newblock \showarticletitle{Strategic coordination of human patrollers and
  mobile sensors with signaling for security games.}. In
  \bibinfo{booktitle}{\emph{AAAI}}. \bibinfo{pages}{1290--1297}.
\newblock


\bibitem[\protect\citeauthoryear{Yin, Jiang, Tambe, Kiekintveld, Leyton-Brown,
  Sandholm, and Sullivan}{Yin et~al\mbox{.}}{2012}]%
        {yin2012trusts}
\bibfield{author}{\bibinfo{person}{Zhengyu Yin}, \bibinfo{person}{Albert~Xin
  Jiang}, \bibinfo{person}{Milind Tambe}, \bibinfo{person}{Christopher
  Kiekintveld}, \bibinfo{person}{Kevin Leyton-Brown}, \bibinfo{person}{Tuomas
  Sandholm}, {and} \bibinfo{person}{John~P Sullivan}.}
  \bibinfo{year}{2012}\natexlab{}.
\newblock \showarticletitle{TRUSTS: Scheduling randomized patrols for fare
  inspection in transit systems using game theory}.
\newblock \bibinfo{journal}{\emph{AI magazine}} \bibinfo{volume}{33},
  \bibinfo{number}{4} (\bibinfo{year}{2012}), \bibinfo{pages}{59--59}.
\newblock


\end{thebibliography}
